\documentclass{article}
\usepackage{fullpage}
\usepackage{amsmath}
\usepackage{amsthm}
\usepackage{amssymb}
\usepackage{url}
\usepackage{graphicx}
\usepackage{verbatim}
\usepackage{url}
\usepackage{float}
\usepackage{pgf,tikz}
\usepackage{mathrsfs}
\usepackage{graphicx}
\usepackage{algorithm2e}
\usetikzlibrary{positioning}
\usetikzlibrary{calc}
\usetikzlibrary{patterns}
\usetikzlibrary{arrows}
\usetikzlibrary{backgrounds}
\usetikzlibrary{shapes.misc}
\usepackage{pgfplots}
\usepackage{pgfplotstable,tabularx}
\usepackage{booktabs}
\usepackage{subfig}
\pgfplotsset{compat=1.13}

\tikzstyle{mybox} = [draw=black, fill=white,  thick,
    rectangle, inner sep=10pt, inner ysep=20pt]
\tikzstyle{mybox} = [draw=black, fill=white,  thick,
    rectangle, inner sep=2pt, inner ysep=2pt]

\newtheorem{thm}{Theorem}

\theoremstyle{definition}

\pgfplotstableread[header=true]{
xlabel dimension triteration trtime maxiter
1 3 31.00 0.02 1 
2 5 23.09 0.03 0 
3 10 33.82 0.10 0 
4 20 155.91 0.43 0 
5 50 220.00 0.67 1 
6 100 506.00 2.37 2 
7 200 7.00 0.06 0 
8 300 417.56 3.88 1 
9 400 3.88 2.21 0 
10 500 1660.57 24.34 2 
}\triangleone

\pgfplotstableread[header=true]{
xlabel dimension triteration trtime smoiteration smotime
1 3 7.4 0.0047 38.7 7.5766
2 5 21.3 0.0734 100.6 20.1609
3 10 85.5 0.6109 124.2 30.5141 
4 50 137.2 1.2031 218.5 80.3563
5 100 167.9 2.4266 256 111.2391
6 200 286.7 6.55 254 126.9531 
7 500 339 14.9031 591.1 288.5109
}\triangleDimensionDataTwo

\pgfplotstableread[header=true]{
xlabel	dimension	triteration	trtime
1	3	26.4	0.0359375
2	5	19.6	0.05625
3	10	8.9	0.028125
4	20	6.25	0.0234375
5	50	11.65	0.075
6	100	14	0.11796875
7	200	9.15	0.13515625
8	300	7.8	0.19296875
9	400	5.15	0.15
10	500	4.75	0.16875
11	1000	3.45	0.21171875
}\triangleSeparation

\pgfplotstableread[header=true]{
xlabel dimension delta triteration	trtime trsparsity trdist	smoiteration smotime smosparsity smodist radius
1 3	2.261790498	206.74	0.169375	4.22	2.3063	35.52	7.3028125	4.68	2.3064	8.160764129
2 5	3.0922345	266.18	0.78375	6.62	3.0895	35.86	8.94875	6.44	3.0895	8.740162833
3 10	4.742867336	349.5	2.626875	10	4.7384	64.58	17.9909375	9.22	4.7379	9.984910453
4 20	6.875885307	366.8	2.8925	13.58	6.8693	156.04	53.029375	14.04	6.8673	11.81835794
5 50	11.1950458	511.06	4.4628125	23.4	11.1843	244.68	110.9378125	26	11.1808	15.34915363
6 100	15.88934397	580.24	8.3303125	31.4	15.874	251.64	134.7734375	35.6	15.8688	19.43160041
7 200	22.58780583	616	14.5571875	44.9	22.5658	306.22	205.886875	50.12	22.5585	25.27170626
8 300	27.52988843	627.94	19.79875	53.6	27.5028	373.14	267.761875	58.7	27.4937	29.66483339
9 400	31.86436088	642.44	25.108125	62.6	31.8332	433.64	302.008125	68.36	31.8226	33.45361996
10 500	35.70479079	638.36	30.325	68.3	35.6698	434.96	311.1425	74.3	35.6568	36.89783326
11 1000	50.28434369	654.8	56.335625	97.52	50.23496078	404.5	358.6215625	108.62	50.21831487	49.85878813
}\triangleDimensionData

\pgfplotstableread[header=true]{
xlabel	dimension	delta	triteration	trtime	trsparsity	trdist	smoiteration	smotime	smosparsity	smodist	radius
1	100	13.94180462	774.4666667	10.98958333	33.4	14.06348879	227.6	129.0697917	37.8	14.05840537	19.35915495
2	100	14.47198882	725.06	10.65625	33.78	14.45793587	250.94	146.0446875	38.56	14.45301165	19.48556798
3	100	15.98964922	578.7	8.333125	31.9	15.9740958	272.16	138.9928125	37.18	15.96849667	19.48556798
4	100	18.85491042	410.1333333	6	29.33333333	18.83650931	256.2	116.765625	33.73333333	18.83045526	19.47749136
5	100	23.63441004	201.6666667	2.973958333	25	23.61185688	320.7333333	118.2947917	30.6	23.60513453	19.47749136
6	100	28.42832549	99.33333333	1.513541667	21.53333333	28.40100844	366	129.875	26.86666667	28.39942501	19.47749136
7	100	33.23500019	48.86666667	0.742708333	19.26666667	33.20396271	353.0666667	113.6114583	25	33.20168623	19.47749136
8	100	52.52853053	13.86666667	0.190625	13.46666667	52.48334235	534.8	126.1979167	17.66666667	52.48616907	19.47749136
9	100	71.88770387	7.866666667	0.114583333	10.8	71.8290749	480.7333333	96.80104167	13.4	71.83875991	19.47749136
}\triangleDistanceData

\pgfplotstableread[header=true]{
xlabel distance triteration trtime smoiteration smotime
1 3.17 26.7 0.02 4.7 2.59
2 6.35 26.1 0.02 8 3.58
3 9.52 43.3 0.04 6.3 2.69 
4 12.69 70.2 0.08 7.1 3.18
5 15.87 69.7 0.06 9.1 3.11 
6 19.05 71.3 0.07 9.6 3.12 
7 25.40 68.4 0.07 14.4 4.07 
8 31.76 65.8 0.06 12.7 4.45 
}\distancethree

\pgfplotstableread[header=true]{
xlabel distance triteration trtime smoiteration smotime
1 3.41 5.2 0.03 3.4 7.92 
2 6.81 4.7 0.04 3.4 3.28 
3 10.22 4.3 0.03 3.1 7.53 
4 13.62 4.2 0.02 3.1 7.39 
5 17.03 4.1 0.04 3.1 7.53 
6 34.05 2.8 0.02 3 3.36 
7 51.08 2 0.05 3 3.15 
8 68.10 2 0.02 3 2.93 
}\distancehundred

\pgfplotstableread[header=true]{
dimension
3
10
50
100
300
500
1K
2K
5K
10K
}\dimensionlabel

\pgfplotstableread[header=true]{
xlabel	dimension	delta	triteration	trtime	trsparsity	trdist	smoiteration	smotime	smosparsity	smodist	radius
1 1000	45.51802571	813.7	5.139583333	113	45.47334076	515.8666667	118.4364583	123.3333333	45.45740418	50.18707072
2 1000	47.56203205	749.1	4.393229167	111.0666667	46.49417082	394.3666667	98.31875	122.0666667	46.47810364	50.20827698
3 1000	50.02858203	654.1	4.480729167	107.2666667	47.65587035	567.3	116.3057292	115.9333333	47.63910141	50.18015443
4 1000	55.06289021	551.2666667	4.3921875	99.9	49.4940813	499.4	97.6640625	110.5333333	49.47706403	50.22258046
5 1000	70.10347041	337.9666667	3.921354167	81.2	53.60221403	752.7333333	116.1291667	90	53.58365962	50.32417348
6 1000	94.96510052	174.8	3.203125	61.66666667	60.48058707	725.5333333	86.67239583	74.4	60.45926668	50.23819602
7 1000	119.8290919	106.8666667	2.873958333	49.43333333	68.94238471	627.0666667	67.0140625	62.2	68.91868722	50.20876571
8 1000	145.0703558	70.23333333	2.744791667	42.26666667	78.44094819	887.1666667	86.690625	53.66666667	78.41392783	50.25697475

}\triangleDistanceData

\pgfplotstableread[header=true]{
xlabel	dimension	triteration	trtime
1	3	102.35	0.68203125
2	10	10.05	0.6375
3	50	26.7	0.6984375
4	100	33	0.778125
5	300	24.85	1.3578125
6	500	14.65	1.8265625
7	1000	9.3	3.40390625
8	2000	3.95	5.7203125
9	5000	4.2	13.68984375
10	10000	4.95	26.5296875
}\triangleSeparation

\pgfplotstableread[header=true]{
xlabel dimension delta triteration	trtime trsparsity trdist smoiteration smotime smosparsity smodist radius
1 3 2.179993064 198.5333333 0.636979167	4.566666667	2.253619071	28.46666667	3.132291667	4.733333333	2.253882333	8.593979133
2 10 4.656819898 374.0666667 1.110416667 10.46666667 3.473299114 70.53333333 6.478125 10 3.472993556 10.39779024
3 50 11.01534796 480.6666667 1.4203125 23.26666667 6.011977008 228.4333333 29.52291667 27.4 6.010684139 15.73594363
4 100 15.80318886 596.3666667 1.766145833 33.43333333 8.476495812 265.9666667 44.02864583 38.43333333 8.473936961 19.89300091
5 300 27.49998731 647.6333333 2.372916667 59.13333333 12.30130924 412.8 85.28958333	65.03333333 12.29750847	30.10708236
6 500 35.36300444 665.1	3.198958333	75.3 16.16059826 493.3666667 90.69010417 82.3 16.15531777 37.11307424
7 1000 50.04143402 678.6666667 4.741666667 107.1 21.01680679 478.2333333 84.66666667 119.4333333 21.00975615	50.20719132
8 2000 70.60219787 674.1666667 5.565625	143.5 27.23214931 613.3	117.075	160.3666667	27.22323013 68.70687938
9 5000 111.4165626 690.5666667 15.4890625 218.3333333 36.60841262 654.6333333 138.7432292 245.2 36.59551458	105.4632061
10 10000 157.1626638 699.3666667 36.3890625	290.0666667	157.0069879	646.766667	165.990625 333.0666667 156.9443726 146.8184307
}\triangleDimensionData
\begin{document}
\definecolor{cqcqcq}{rgb}{0.7529411764705882,0.7529411764705882,0.7529411764705882}
\definecolor{zzttqq}{rgb}{0.6,0.2,0.}
\definecolor{qqqqff}{rgb}{0.,0.,1.}

%%%%%%%%% TITLE
\title{A Comparison of the Triangle Algorithm and SMO for Solving the Hard Margin Problem}

\author{Mayank Gupta \qquad
%Rutgers University\\
%{\tt\small mayank.gupta.cs@rutgers.edu}
Bahman Kalantari\\
Rutgers University \\
{\tt\small mayank.gupta.cs@rutgers.edu, kalantari@cs.rutgers.edu}
%{\small\url{kalantari@cs.rutgers.edu}}
}
\providecommand{\keywords}[1]{\textbf{\textit{Keywords:}} #1}
\maketitle
% \thispagestyle{empty}

%%%%%%%%% ABSTRACT
\begin{abstract}
In this article we consider the problem of testing, for two finite sets of points in the Euclidean space, if their convex hulls are disjoint and computing an optimal supporting hyperplane if so. This is a fundamental problem of classification in machine learning known as the {\it hard-margin SVM}.
The problem can be formulated as a quadratic programming problem. The {\it SMO algorithm}~\cite{plattsmo} is the current state of art algorithm for solving it, but it does not answer the question of separability. 
An alternative to solving both problems is the {\it Triangle Algorithm}~\cite{trianglesvm}, a geometrically inspired algorithm, initially described for the {\it convex hull membership} problem~\cite{triangle}, a fundamental problem in linear programming.
First, we describe the experimental performance of the Triangle Algorithm for testing the intersection of two convex hulls. Next, we compare the performance of Triangle Algorithm with SMO for finding the optimal supporting hyperplane.
Based on experimental results ranging up to 5000 points in each set in dimensions up to $10000$, the Triangle Algorithm outperforms SMO.
\end{abstract}
%%%%%%%%% BODY TEXT
\section{Introduction}
Given a pair of finite sets, determining if they are linearly separable and if so, finding a separating hyperplane is a problem of classification dealt with in statistics and machine learning.
In two-class classification, we wish to estimate a function $f : \mathbb{R}^m \rightarrow \{\pm1\}$ using the input-output data
\begin{equation}
    (x_1,y_1),\dots , (x_n,y_n) \in \mathbb{R}^m \times \{\pm1\}
\end{equation}
Given some new data point $x$, we use $f(x)$ to classify its label.
According to Vapnik-Chervonenkis theory \cite{vctheory} minimizing the error on the test set depends not only on minimizing the empirical risk but also on the capacity of the function class. This has led to the development of the class of functions whose capacity can be computed.

Vapnik and Chervonenkis  \cite{vctheory} and Vapnik and Lerner \cite{VapLer63} considered the class of hyperplanes
\begin{equation}\label{eq:sepHyperplane}
    w^Tx - b = 0, \quad w \in \mathbb{R}^m, b \in \mathbb{R}
\end{equation}
corresponding to the decision function
\begin{equation}\label{eq:decisionFunc}
    f(x) = sign( w^Tx - b),
\end{equation}
and proposed a learning algorithm for separable problems, to construct the function $f$ from empirical data.
This optimal margin classifier~\cite{vapnik79} for the linearly separable case is based upon the minimization of the number of classification errors by placing optimal boundaries between classes. This has also led to the tremendous success of {\it support vector machines} (SVM) in classification tasks.

\begin{figure}[H]
    \centering
\definecolor{uuuuuu}{rgb}{0.26666666666666666,0.26666666666666666,0.26666666666666666}
\begin{tikzpicture}[line cap=round,line join=round,>=triangle 45,x=1.0cm,y=1.0cm,scale=0.5]

\clip(-0.42,-0.36) rectangle (16.,12.);
\draw [dash pattern=on 3pt off 3pt,domain=-0.42:16.] plot(\x,{(--21.84-2.52*\x)/1.68});
\draw [dash pattern=on 3pt off 3pt,domain=-0.42:16.] plot(\x,{(--36.0864-2.52*\x)/1.68});
\draw [domain=-0.42:16.] plot(\x,{(--29.0976-2.52*\x)/1.68});
\draw (7.34,11.44) node[anchor=north west] {$\mathit{\mathbf{V'}}$};
\draw (0.04,7.74) node[anchor=north west] {$\mathit{\mathbf{V}}$};
\draw [line width=0.4pt] (4.32,6.52)-- (2.38,6.7);
\draw [line width=0.4pt] (2.38,6.7)-- (0.92,3.18);
\draw [line width=0.4pt] (0.92,3.18)-- (2.8,2.26);
\draw [line width=0.4pt] (2.8,2.26)-- (4.8,1.66);
\draw [line width=0.4pt] (4.8,1.66)-- (6.22,2.52);
\draw [line width=0.4pt] (6.22,2.52)-- (6.,4.);
\draw [line width=0.4pt] (6.,4.)-- (4.32,6.52);
\draw [line width=0.4pt] (8.9,10.42)-- (9.32,7.5);
\draw [line width=0.4pt] (9.32,7.5)-- (12.96,4.3);
\draw [line width=0.4pt] (12.96,4.3)-- (14.78,8.48);
\draw [line width=0.4pt] (14.78,8.48)-- (12.58,9.46);
\draw [line width=0.4pt] (12.58,9.46)-- (8.9,10.42);
\draw [line width=0.4pt] (6.313846153846154,7.8492307692307675)-- (4.32,6.52);
\draw [line width=0.4pt] (5.242038550432671,7.134692366955111) -- (5.250359053376049,7.284461419935924);
\draw [line width=0.4pt] (5.242038550432671,7.134692366955111) -- (5.383487100470105,7.08476934929484);
\draw [line width=0.4pt] (6.9169230769230765,6.944615384615385)-- (4.,5.);
\draw [line width=0.4pt] (5.3835770119711315,5.9223846746474225) -- (5.39189751491451,6.072153727628234);
\draw [line width=0.4pt] (5.3835770119711315,5.9223846746474225) -- (5.525025562008565,5.872461656987151);
\draw (7.4,6.22)-- (9.32,7.5);
\draw (9.68,3.24) node[anchor=north west] {$\mathit{\mathbf{H}}$};
\begin{scriptsize}
\draw [fill=black] (2.48,4.74) circle (2.5pt);
\draw [fill=black] (2.8,2.26) circle (2.5pt);
\draw [fill=black] (0.92,3.18) circle (2.5pt);
\draw [fill=black] (4.8,1.66) circle (2.5pt);
\draw [fill=black] (4.,5.) circle (2.5pt);
\draw[color=black] (4.18,5.42) node {$v_3$};
\draw [fill=black] (4.8,3.14) circle (2.5pt);
\draw [fill=black] (6.22,2.52) circle (2.5pt);
\draw [fill=black] (4.32,6.52) circle (2.5pt);
\draw[color=black] (4.5,6.94) node {$v_1$};
\draw [fill=black] (6.,4.) circle (2.5pt);
\draw[color=black] (6.18,4.42) node {$v_2$};
\draw [fill=black] (2.38,6.7) circle (2.5pt);
\draw[color=black] (-0.26,13.14) node {$a$};
\draw [color=black] (9.32,7.5) circle (2.5pt);
\draw[color=black] (9.56,7.92) node {$v_1'$};
\draw[color=black] (2.98,17.38) node {$b$};
\draw [color=black] (10.56,9.8) circle (2.5pt);
\draw [color=black] (12.44,6.82) circle (2.5pt);
\draw [color=black] (13.36,5.4) circle (2.5pt);
\draw [color=black] (14.78,8.48) circle (2.5pt);
\draw [color=black] (10.94,8.5) circle (2.5pt);
\draw [color=black] (8.9,10.42) circle (2.5pt);
\draw [color=black] (9.72,8.86) circle (2.5pt);
\draw [color=black] (12.58,9.46) circle (2.5pt);
\draw [color=black] (12.96,4.3) circle (2.5pt);
\draw[color=black] (0.2,17.38) node {$c$};
\draw [fill=uuuuuu] (6.313846153846154,7.8492307692307675) circle (1.5pt);
\draw[color=uuuuuu] (6.46,8.12) node {$I$};
\draw[color=black] (5.52,7.68) node {$\gamma_1$};
\draw[color=black] (5.68,6.46) node {$\gamma_3$};
\draw [fill=uuuuuu] (7.4,6.22) circle (1.5pt);
\draw[color=uuuuuu] (7.54,6.5) node {$F$};
\draw[color=black] (8.98,6.82) node {$\gamma_1'$};
\end{scriptsize}
\end{tikzpicture}
    \caption{Binary classification}
    \label{fig:bclass}
\end{figure}
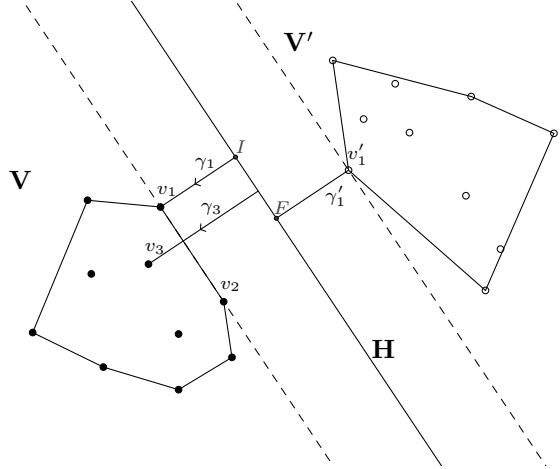

Figure \ref{fig:bclass} is an example of a binary or two-class classification problem. Here $V=\{v_1,\dots , v_n\}$ and $V' = \{v_1',\dots , v_{n'}'\}$ are samples from the two classes. The convex hulls, $K=conv(V)$ and $K'=conv(V')$, for both classes are drawn. The class membership $y_i$ is 1 if $x \in V'$ and -1 if $x \in V$.

According to the classical separating hyperplane theorem, $K$ and $K'$ are disjoint if and only if they can
be separated by a hyperplane, i.e. if there exists $w \in \mathbb{R}^m$ and $b \in \mathbb{R}$ such that
\begin{equation}
    \begin{split}
        w^Tx < b,& \quad \forall x \in K \\
        w^Tx > b,& \quad \forall x \in K'
    \end{split}
\end{equation}

The hyperplane
\begin{equation}
    W = \{ x \in \mathbb{R}^m : \quad w^Tx = b\}
\end{equation}
separates $K$ and $K'$.

The optimal hyperplane $H$ is orthogonal to the shortest line connecting the convex hulls, $K$ and $K'$, of the two classes, and intersects this line segment half-way. Among all the hyperplanes separating the data, $H$ is the unique one yielding the maximum margin of separation between $K$ and $K'$.
The distance between a test point and the optimal hyperplane also provides an estimate on the confidence in the classification.

In Figure \ref{fig:bclass}, the minimum distance $\gamma_1$ of K to the hyperplane is given by the line segment $Iv_1$. The point $v_1$ represents $x^{(i)}$, and the point $I$ can be represented as
\begin{equation}
    I = v_1 - \gamma_1 \times \frac{w}{||w||}
\end{equation}
Since $I$ lies on the hyperplane $w^Tx -b = 0$
\begin{equation}
    \begin{split}
         w^T I &-b = 0 \\
         w^T(x^{(i)} - {\gamma_i}.\frac{w}{||w||}) &- b = 0 
    \end{split}
\end{equation}

Solving for $\gamma_i$ lends
\begin{equation} \label{eq:gammasolve}
    \gamma_i = y^{(i)}(\frac{w}{||w||})^T x^{(i)} - \frac{b}{||w||})
\end{equation}

To obtain a unique solution, H, to (\ref{eq:gammasolve}), we set $||w|| = 1$. By definition the maximally separating hyperplane will be such that
\begin{equation} \label{eq:gammamin}
    \gamma = min\{ \gamma_i : i=1,\dots,m\}
\end{equation}

The solution to this relies on the fact that the minimum of the Euclidean distance function $d(x,x')$ is attained and is positive
\begin{equation}
    min \{ d(x,x') : x \in K, x' \in K' \} > 0
\end{equation}

To find the maximum of the minimum margins, (\ref{eq:gammamin}) can be written as
\begin{equation}
\begin{split}
    & \underset{\gamma, w, b}{max} \quad \gamma\\
    s.t.\quad & y_i (w^Tx^{(i)} + b) \ge \gamma, \quad i=1,\dots,m \\
    & ||w|| = 1
\end{split}
\end{equation}

Our goal in this article is to consider the hard margin problem and study the performance of two distinct algorithms; the  {\it Triangle Algorithm (TA)} and the {\it Sequential Minimal Optimization(SMO)} algorithm.
The Triangle Algorithm is a geometrically inspired algorithm while the SMO algorithm is the state of the art quadratic programming method.

There exist another geometric algorithm, Gilbert's algorithm \cite{gilbert},
based on the Frank Wolfe approximation\cite{frankwolfe} for quadratic programs,
that were empirically shown to converge fast\cite{keerthi}. 
It wasn't until 2009 that Gartner et al \cite{jaggi} gave the proof on their convergence speed. They however do not consider the case if the convex hulls are non-separable. 
The Triangle Algorithm belongs to the same class of algorithm but also offer convergence bounds for answering the question of separability.
Earlier results\cite{triangleperformance}, for the special case of a point and a set, show that the Triangle Algorithm outperforms the Frank Wolfe algorithm.
%Our goal in this article is to study the performance of the {\it Triangle Algorithm} to test the intersection of two finite convex hulls and also compare it to the {\it Sequential Minimal Optimization(SMO)} algorithm for calculating the optimal hyperplane in the case of hard margin problem.

The Triangle Algorithm thus works in two phases. In the first phase, given $(K,K')$ described above, it determines whether they are separable or not. The first phase begins with a pair of iterates $(p,p') \in K \times K'$ .In each iteration it either moves $p$ close to $p'$ using a {\it $p$-pivot} to a new point $p \in K$, or it moves $p'$ closer to $p$ using a {\it $p'$-pivot}. A $p$-pivot is a point  $v \in V$ that lies in the {\it Voronoi region} (\ref{eq:Voronoi}) of $p'$, $Vor(p')$.
\begin{equation} \label{eq:Voronoi}
Vor(p') = \{ x \in \mathbb{R}^m:, d(x,p') < d(x,p)\}
\end{equation}
Similarly for $p'$-pivot. If the Triangle Algorithm fails to find any such pivot, it implies that $K \in Vor(p)$ and $K \in Vor(p')$, returning a {\it witness pair} $(p,p') \in K \times K'$, the perpendicular bisector of which is a hyperplane that separates $(K,K')$. In the second phase it gradually reduces the gap between $p$ and $p'$ to compute the optimal hyperplane.

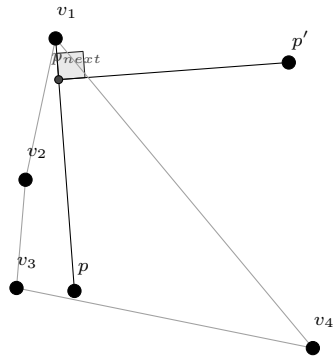
\begin{figure}[H]
\centering
\definecolor{aqaqaq}{rgb}{0.6274509803921569,0.6274509803921569,0.6274509803921569}
\definecolor{qqwuqq}{rgb}{0.12941176470588237,0.12941176470588237,0.12941176470588237}
\definecolor{uuuuuu}{rgb}{0.26666666666666666,0.26666666666666666,0.26666666666666666}
\begin{tikzpicture}[line cap=round,line join=round,>=triangle 45,x=1.0cm,y=1.0cm]
\clip(0.5,0.5) rectangle (5.5,6.5);
\draw[color=qqwuqq,fill=qqwuqq,fill opacity=0.1] (2.1105445240742347,4.617917224170841) -- (2.0844653664109085,4.967577487142321) -- (1.7348051034394292,4.9414983294789945) -- (1.7608842611027553,4.591838066507515) -- cycle; 
\draw (1.72,5.14)-- (1.9705785123966948,1.780330578512403);
\draw (4.82,4.82)-- (1.7608842611027553,4.591838066507515);
\draw [color=aqaqaq] (1.72,5.14)-- (1.32,3.26);
\draw [color=aqaqaq] (1.32,3.26)-- (1.2,1.82);
\draw [color=aqaqaq] (1.2,1.82)-- (5.14,1.02);
\draw [color=aqaqaq] (5.14,1.02)-- (1.72,5.14);
\begin{scriptsize}
\draw [fill=black] (1.72,5.14) circle (2.5pt);
\draw[color=black] (1.8714049586776864,5.482809917355378) node {$v_1$};
\draw [fill=black] (1.32,3.26) circle (2.5pt);
\draw[color=black] (1.4747107438016533,3.6150413223140556) node {$v_2$};
\draw [fill=black] (4.82,4.82) circle (2.5pt);
\draw[color=black] (4.978842975206612,5.1191735537190155) node {$p'$};
\draw [fill=black] (1.9705785123966948,1.780330578512403) circle (2.5pt);
\draw[color=black] (2.0862809917355376,2.0778512396694278) node {$p$};
\draw [fill=black] (1.2,1.82) circle (2.5pt);
\draw[color=black] (1.3424793388429757,2.1604958677686015) node {$v_3$};
\draw [fill=black] (5.14,1.02) circle (2.5pt);
\draw[color=black] (5.2928925619834715,1.3671074380165353) node {$v_4$};
\draw [fill=uuuuuu] (1.7608842611027553,4.591838066507515) circle (1.5pt);
\draw[color=uuuuuu] (2.003636363636364,4.8712396694214934) node {$p_{next}$};
\end{scriptsize}
\end{tikzpicture}
    \caption{Triangle Algorithm: Moving $p$ closer to $p'$}
    \label{fig:trpivot}
\end{figure}

Figure \ref{fig:trpivot} shows $p$ moving to the point closest to $p'$ along the segment $pv_1$, such that $d(p_{next},p') < d(p,p')$. The convex hull $K=conv(\{v_1,v_2,v_3,v_4\})$ is also drawn.

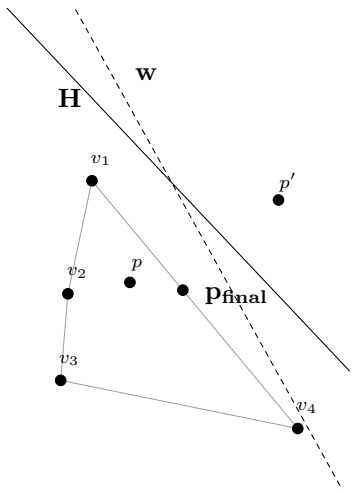
\begin{figure}[H]
    \centering
    \definecolor{aqaqaq}{rgb}{0.6274509803921569,0.6274509803921569,0.6274509803921569}
\begin{tikzpicture}[line cap=round,line join=round,>=triangle 45,x=1.0cm,y=1.0cm,scale=0.8]
\clip(0.,0.) rectangle (6.,8.);
\draw [color=aqaqaq] (1.72,5.14)-- (1.32,3.26);
\draw [color=aqaqaq] (1.32,3.26)-- (1.2,1.82);
\draw [color=aqaqaq] (1.2,1.82)-- (5.14,1.02);
\draw [color=aqaqaq] (5.14,1.02)-- (1.72,5.14);
\draw [dash pattern=on 2pt off 2pt,domain=0.:6.] plot(\x,{(-14.519007130660455--2.4692561983471073*\x)/-1.3702479338842917});
\draw [domain=0.:6.] plot(\x,{(-12.501923045849473--1.5896365732544226*\x)/-1.4995021398221584});
\draw (3.441652892561984,3.5158677685950477) node[anchor=north west] {$\mathbf{p_{final}}$};
\draw (2.2846280991735544,7.168760330578519) node[anchor=north west] {$\mathbf{w}$};
\draw (1.0,6.82165289256199) node[anchor=north west] {$\mathbf{H}$};
\begin{scriptsize}
\draw [fill=black] (1.72,5.14) circle (2.5pt);
\draw[color=black] (1.8714049586776864,5.482809917355378) node {$v_1$};
\draw [fill=black] (1.32,3.26) circle (2.5pt);
\draw[color=black] (1.4747107438016533,3.6150413223140556) node {$v_2$};
\draw [fill=black] (4.82,4.82) circle (2.5pt);
\draw[color=black] (4.978842975206612,5.1191735537190155) node {$p'$};
\draw [fill=black] (2.350743801652893,3.4497520661157086) circle (2.5pt);
\draw[color=black] (2.466446280991736,3.7472727272727337) node {$p$};
\draw [fill=black] (1.2,1.82) circle (2.5pt);
\draw[color=black] (1.3424793388429757,2.1604958677686015) node {$v_3$};
\draw [fill=black] (5.14,1.02) circle (2.5pt);
\draw[color=black] (5.2928925619834715,1.3671074380165353) node {$v_4$};
\draw[color=black] (-0.5418181818181813,11.367107438016536) node {$a$};
\draw [fill=black] (3.2303634267455776,3.320497860177842) circle (2.5pt);
\draw[color=black] (-0.5418181818181813,8.75553719008265) node {$b$};
\end{scriptsize}
\end{tikzpicture}
    \caption{ $(p, p')$ are witness pairs in Triangle Algorithm II}
    \label{fig:tatwo}
\end{figure}

In the next phase the Triangle Algorithm starts with a witness pair $(p,p')$ and its orthogonal separating hyperplane $w^Tx=b$ to calculate the optimal hyperplane $H$. The distance, $d(p,p')$, is an upper bound to the optimal distance between $K$ and $K'$. It computes a lower bound to the distance and reduces $d(p,p')$ by moving $p$ to $p_{final}$. It does so by finding a point nearest to $p'$ on the segment $pv_{ext}$, where $v_{ext}$ is an {\it extreme vertex} of $K$ (see figure \ref{fig:tatwo}) defined as

\begin{equation}
    v_{ext} = max\{w^Tv_i: \quad v_i \in V\}
\end{equation}
In this case $v_{ext}$ is not a pivot, yet it allows reducing $d(p,p')$. It is called a {\it weak pivot}. In the next iteration it uses either a pivot or a weak pivot.
The algorithm terminates when the lower bound and upper bound are close within a prescribed tolerance.
Figure \ref{fig:tatwo} shows a special case of a set and a point for Triangle Algorithm II. The input parameters are $(p,p'), V$ and the output is $(p_{final},p')$. Here $v_1$ and $v_4$ act as the extreme points.

The SMO algorithm however, is a quadratic programming method for solving the problem posed by (\ref{eq:gammamin}). Setting the derivative of its Lagrangian to zero (see section \ref{smosection} for details), we get
\begin{equation} \label{eq:smow}
    w = \sum_{i=1}^n \alpha_i y^{(i)} x^{(i)}
\end{equation}
\begin{equation}\label{eq:smokkt1}
    \sum_{i=1}^n \alpha_i y^{(i)} = 0
\end{equation}
as a partial set of equations.
The key idea of SMO is to fix all $\alpha_i$'s except for a pair$(\alpha_i,\alpha_j)$ of them. Having selected $(\alpha_i,\alpha_j)$, SMO then reoptimizes $w$ with respect to them.
%This represents the optimal hyperplane $w$ as a linear combination of the input vectors.
%SMO performs a coordinate descent over $\alpha$ to minimize $w$. Due to (\ref{eq:smokkt1}) we cannot reduce the error by picking one of the $\alpha_i$s. The SMO algorithm therefore adjusts a pair $(\alpha_1, \alpha_2)$ such that the following constraint is respected along with the other KKT conditions
%\begin{equation} \label{eq:smoa1a2}
%    y^{(1)}\alpha_1 + y^{(2)}\alpha_2 = -\sum_{i=3}^n \alpha_i y^{(i)}
%\end{equation}
%
%The SMO algorithm can be described as following:
%\begin{itemize}
 %   \item Select a pair $(\alpha_i,\alpha_j)$ to update.
  %  \item Reoptimize $W(\alpha)$ with respect to $(\alpha_i,\alpha_j)$, while holding other $\alpha_k$ ($k \ne i,j$) fixed.
%\end{itemize}

The SMO algorithm, as we shall see in section \ref{smosection}, does not answer the question of separability of the two sets.

For the intended comparison we implemented both algorithms, triangle algorithm and SMO, in matlab. For our experiments we generated two unit balls with random mean and placed them at some distance apart, to analyze the results of both algorithms. We tested and analyzed the results for both algorithms for up to 1000 dimensions with up to 2000 points in each set. 

The remainder of this article is organized as follows. In section \ref{trsection} we describe the {\it distance duality}, Triangle Algorithm and its complexity. In section \ref{smosection} we describe the {\it langrange duality} and SMO algorithm. In section \ref{section:intersection}, we describe the performance of the Triangle Algorithm for testing the intersection or separation of two convex hulls. In section \ref{section:optimal}, we compare the performance of the Triangle Algorithm with SMO for finding the optimal hyperplane. In section \ref{section:implementation}, we discuss some ideas for an efficient implementation of the Triangle Algorithm and future work.

\section{The Triangle Algorithm} \label{trsection}

Given a finite set $V= \{v_1,\dots,v_n\} \subset \mathbb{R}^m$, and a distinguished point $p \in \mathbb{R}^m$, the \textit{convex hull membership problem}(or \textit{convex hull decision problem}), is to test if $p \in conv(V)$, or convex hull of V.
Given a desired tolerance $\epsilon \in (0,1)$, we call a point $p_\epsilon \in conv(V)$ an $\epsilon$-approximate solution if $d(p_\epsilon,p) \le \epsilon R$, where $R = max\{d(p,v_i): \quad i=1,\dots,n\}$.

A recent algorithm for the convex hull membership problem is the Triangle Algorithm~\cite{triangle}. It can either
compute an $\epsilon$-approximate solution, or when $p \notin conv(V)$ a separating hyperplane and a point that approximates
the distance from $p$ to $conv(V)$ to within a factor of 2. Based on preliminary experiments, the
Triangle Algorithm performs quite well on reasonably large size problem, see ~\cite{triangleperformance}.

Here we review the terminology and some results from ~\cite{triangle}~\cite{trianglesvm} for the case of two finite convex hulls. 

\subsection{Definitions}
The Euclidean distance is denoted by $d(\cdot, \cdot)$.
Let $V=\{v_1, \dots, v_n\}$ and $V'=\{v_1', \dots, v_{n'}'\}$,  $K= conv(V)$ and $K'= conv(V')$.
Assume we are given $p_0 \in K$, $p'_0 \in K'$.
Let
\begin{equation} \label{eqa4}
\delta_*=d(K,K')= \min \{d(p,p'): p \in K, p' \in K' \}.
\end{equation}
It is easy to prove that $\delta_* = 0$ if and only if $K \cap K' \not = \emptyset$.
Suppose $\delta_*=0$. We say a pair $(p,p') \in K \times K'$ is an $\epsilon$-{\it approximation solution to the intersection problem} if
\begin{equation}
d(p,p') \leq \epsilon d(p,v), \quad \text{for some} \quad  v \in K, \quad \text{or} \quad  d(p,p') \leq \epsilon d(p',v'), \quad \text{for some} \quad v' \in K'.
\end{equation}

Triangle Algorithm I computes  an $\epsilon$-approximate solution to the intersection problem when $\delta_*=0$, or a pair of separating hyperplane when $\delta_* >0$.
In each iteration it begins with a pair of iterates $p_i \in K$, $p_i' \in K'$ and searches for a $p'$-pivot $v \in V$ and $p$-pivot $v' \in V'$ along with the optimal pair of supporting hyperplanes defined below. 

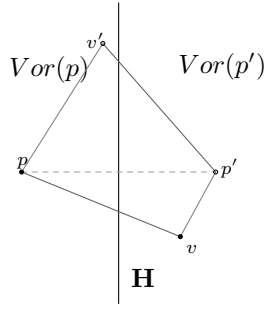
\begin{figure}[H]
\centering
\definecolor{yqyqyq}{rgb}{0.5019607843137255,0.5019607843137255,0.5019607843137255}
\definecolor{aqaqaq}{rgb}{0.6274509803921569,0.6274509803921569,0.6274509803921569}
\definecolor{wqwqwq}{rgb}{0.3764705882352941,0.3764705882352941,0.3764705882352941}
\begin{tikzpicture}[line cap=round,line join=round,>=triangle 45,x=1.0cm,y=1.0cm,scale=0.5]
\clip(1.5,1.) rectangle (9.,9.);
\draw (5.16,1.) -- (5.16,9.);
\draw (6.5,7.9146984375331515) node[anchor=north west] {$Vor(p')$};
\draw (2.0,7.837091433225092) node[anchor=north west] {$Vor(p)$};
\draw [color=wqwqwq] (2.58,4.5)-- (6.8,2.78);
\draw [color=wqwqwq] (7.74,4.5)-- (4.74,7.92);
\draw (5.243424137857897,2.171780118736762) node[anchor=north west] {$\mathbf{H}$};
\draw [dash pattern=on 2pt off 2pt,color=aqaqaq] (2.58,4.5)-- (7.74,4.5);
\draw [line width=0.4pt,color=yqyqyq] (2.58,4.5)-- (4.74,7.92);
\draw [line width=0.4pt,color=yqyqyq] (6.8,2.78)-- (7.74,4.5);
\begin{scriptsize}
\draw [fill=black] (2.58,4.5) circle (1.5pt);
\draw[color=black] (2.6047859913838827,4.73281126090272) node {$p$};
\draw [color=black] (7.74,4.5) circle (1.5pt);
\draw[color=black] (8.11488329725609,4.596999003363615) node {$p'$};
\draw [color=black] (4.74,7.92) circle (1.5pt);
\draw[color=black] (4.544961099085364,7.992305441841211) node {$v'$};
\draw [fill=black] (6.8,2.78) circle (1.5pt);
\draw[color=black] (7.105992241251319,2.462806384891984) node {$v$};
\end{scriptsize}
\end{tikzpicture}
\caption{$v$ is $p'$-pivot for $p$ (left); $v'$ is $p$-pivot for $p'$}
\label{fig:pivot}
\end{figure}

Given a pair  $(p,p') \in K \times K'$ (see Figure \ref{fig:pivot}), we say $v \in K$ is a $p'$-{\it pivot} for $p$ if
\begin{equation} \label{def21}
d(p,v) \geq d(p',v).
\end{equation}
We say $v' \in K'$ is a $p$-{\it pivot} for $p'$ if
\begin{equation} \label{def22}
d(p',v') \geq d(p,v').
\end{equation}

Consider the Voronoi diagram of the set $\{p, p' \}$ and the corresponding  Voronoi cells
\begin{equation}
Vor(p)= \{x: d(x,p) < d(x,p')\}, \quad  Vor(p')= \{x: d(x,p') < d(x,p)\}.
\end{equation}
If $H=\{x: h^Tx= a \}$ is the orthogonal bisecting hyperplane of the line $pp'$, it intersects  $K$ if and only if there exists $v \in K$ that is a $p'$-pivot for $p$, and $H$ intersects  $K'$ if and only if there exists
$v' \in K'$ that is a $p$-pivot for $p'$. Given $(p,p') \in K \times K'$, we say it is a \emph{witness pair} if the orthogonal bisecting hyperplane of the line segment $pp'$ separates $K$ and $K'$. Figure \ref{fig:tatwo} shows an example of a witness-pair $(p,p')$ with $w$ as their orthogonal bisecting hyperplane.
In Figure \ref{fig:pivot}, the point $v$ and $v'$ are pivots for $p'$ and $p$, respectively. The four points $p,p',v,v'$ need not be coplanar.

Each iteration of Triangle Algorithm I requires computing for a given pair $(p,p') \in K \times K'$ a $p'$-pivot $v$ for $p$, or a $p$-pivot $v'$ for $p'$. By squaring (\ref{def21}) and (\ref{def22}), these are respectively equivalent to checking if

\begin{equation} \label{pivots}
2v^T(p'-p) \geq \Vert p' \Vert^2 -  \Vert p \Vert^2, \quad 2v'^T(p-p') \geq \Vert p \Vert^2 -  \Vert p' \Vert^2.
\end{equation}

From the above it follows that the existence and computation of a pivot can be carried out by solving the following
\begin{equation} \label{eqa5}
\max \{(p'-p)^Tv:  \quad v \in V\}, \quad \max \{(p-p')^Tv': \quad  v' \in V' \}.
\end{equation}

Triangle Algorithm II begins with a witness pair $(p, p') \in K \times K'$, then it computes an $\epsilon$-approximate solution to the distance problem.  Since $(p, p')$ is a witness pair, there exists no $p'$-pivot for $p$, or a $p$-pivot for $p'$. 

We say a witness pair $(p,p') \in K \times K'$ is an $\epsilon$-{\it approximation solution to the distance problem} (or $\epsilon$-{\it approximation solution to} $\delta_*$) if
\begin{equation}
d(p,p') - \delta_* \leq \epsilon d(p,p').
\end{equation}
A pair of  parallel hyperplanes $(H,H')$ {\it supports} $(K,K')$, if $H$ contains a boundary point of $K$, $H'$ contains a boundary point of $K'$,
$K \subset H_+$, $K' \subset H'_+$, where $H_+, H'_+$  are disjoint halfspaces corresponding to $H,H'$.
Figure \ref{fig:bclass} shows an example of hyperplanes supported by boundary points.
A witness pair $(p,p') \in K \times K'$ is an $\epsilon$-{\it approximate solution to the supporting hyperplanes problem} if
\begin{equation}
    d(p,p') - \delta_* \leq \epsilon d(p,p')
\end{equation}
and there exists a pair of parallel supporting hyperplanes $(H, H')$  orthogonal to the line segment $pp'$ such that the distance between them satisfies
\begin{equation}
    \delta_* - d(H,H') \leq \epsilon d(p,p')
\end{equation}

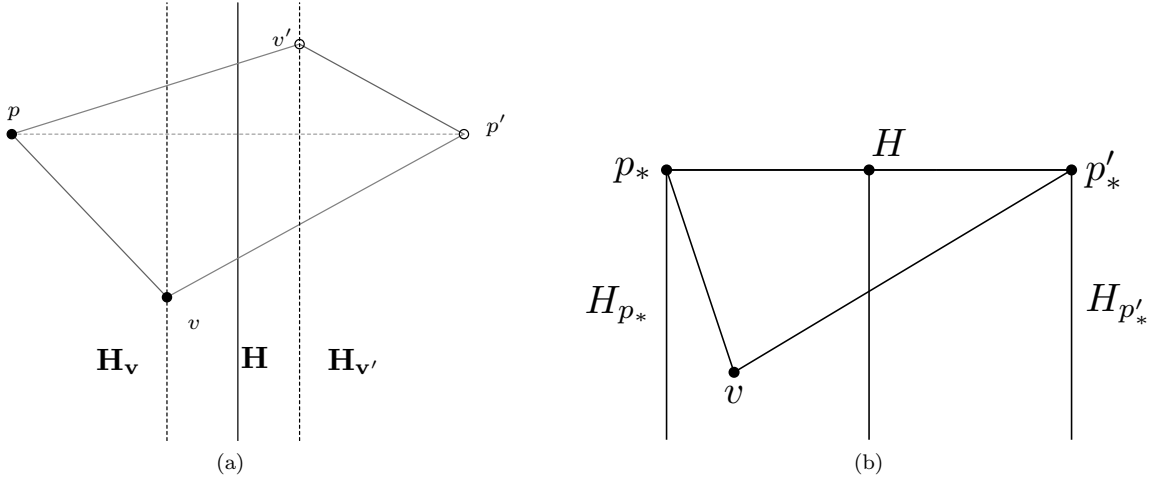
\begin{figure}[H]
\subfloat[]{{
\resizebox {0.5\columnwidth} {!} {
\definecolor{yqyqyq}{rgb}{0.5019607843137255,0.5019607843137255,0.5019607843137255}
\definecolor{aqaqaq}{rgb}{0.6274509803921569,0.6274509803921569,0.6274509803921569}
\definecolor{wqwqwq}{rgb}{0.3764705882352941,0.3764705882352941,0.3764705882352941}
\begin{tikzpicture}[line cap=round,line join=round,>=triangle 45,x=1.0cm,y=1.0cm]
\clip(2.,1.) rectangle (8.5,6.);
\draw (5.16,1.) -- (5.16,6.);
\draw [color=wqwqwq] (2.58,4.5)-- (4.350943588315216,2.637422144585118);
\draw [color=wqwqwq] (7.74,4.5)-- (5.864280172322371,5.528283055060328);
\draw (5.068808378164763,2.191181869813777) node[anchor=north west] {$\mathbf{H}$};
\draw [dash pattern=on 1pt off 1pt,color=aqaqaq] (2.58,4.5)-- (7.74,4.5);
\draw [line width=0.4pt,color=yqyqyq] (2.58,4.5)-- (5.864280172322371,5.528283055060328);
\draw [line width=0.4pt,color=yqyqyq] (4.350943588315216,2.637422144585118)-- (7.74,4.5);
\draw [dash pattern=on 1pt off 1pt] (5.864280172322371,1.) -- (5.864280172322371,6.);
\draw [dash pattern=on 1pt off 1pt] (4.350943588315216,1.) -- (4.350943588315216,6.);
\draw (6.058297683092519,2.171780118736762) node[anchor=north west] {$\mathbf{H_{v'}}$};
\draw (3.4196595366185045,2.171780118736762) node[anchor=north west] {$\mathbf{H_v}$};
\begin{scriptsize}
\draw [fill=black] (2.58,4.5) circle (1.5pt);
\draw[color=black] (2.6047859913838827,4.73281126090272) node {$p$};
\draw [color=black] (7.74,4.5) circle (1.5pt);
\draw[color=black] (8.11488329725609,4.596999003363615) node {$p'$};
\draw [color=black] (5.864280172322371,5.528283055060328) circle (1.5pt);
\draw[color=black] (5.670262661552223,5.605890059368387) node {$v'$};
\draw [fill=black] (4.350943588315216,2.637422144585118) circle (1.5pt);
\draw[color=black] (4.661371605547453,2.3269941273528802) node {$v$};
\end{scriptsize}
\end{tikzpicture}
%\caption{The point $v$ is a weak $p'$-pivot for $p$, but not a $p'$-pivot.}
\label{fig:weakpivot}
}
}}
\subfloat[]{{
\resizebox {0.5\columnwidth} {!} {
\begin{tikzpicture}[scale=0.6]
\draw (-4,0) -- (2,0) node[pos=0.55, above] {$H$};
\draw (0,0) -- (2,0) node[pos=0.5, above] {};
\draw (-4,0) -- (-1,0) node[pos=0.5, above] {};
\draw (-1,0) -- (0,0) node[pos=0.55, above] {};
\filldraw (-3,-3) circle (2pt);
\draw (-3,-3) node[below] {$v$};
\draw (-4,0) -- (-4,-4) node[pos=0.5, left] {$H_{p_*}$};
\draw (2,0) -- (2,-4) node[pos=0.5, right] {$H_{p'_*}$};
\filldraw (-4,0) circle (2pt);
\filldraw (2,0) circle (2pt);
\draw (-4,0) node[left] {$p_*$};
\draw (2,0) node[right] {$p_*'$};
\draw (-1,0) node[above] {};
\filldraw (-1,0) circle (2pt);
\draw (-3,-3.) -- (2,0) node[pos=0.3, left] {};
\draw (-3,-3.) -- (-4,0) node[pos=0.8, left] {};
\draw (-1,0) -- (-1,-4) node[pos=0.5] {};
\end{tikzpicture}
%\caption{If $H_{p_*}$ is not supporting $K$, $d(p_*,p'_*)$ is not optimal.} 
\label{Fig89}
}}
}
\caption{Hyperplanes depicting the lower and upper bounds on the optimal distance}
\end{figure}

If  $(p, p')$ is not already an $\epsilon$-approximate solution to $\delta_*$, the algorithm makes use of a {\it weak-pivot}.
Given a witness pair  $(p,p') \in K \times K'$,  suppose that $H$ is the orthogonal bisecting hyperplane of the line segment $pp'$. We shall say
$v \in V$ is a {\it weak-pivot} $p'$ for $p$ if
\begin{equation}
d(p, H) >  d(v, H)
\end{equation}
(i.e. if $H_v$ is the hyperplane parallel to $H$ passing through $v$, it separates $p$ from $p'$, see Figure \ref{fig:weakpivot}).  Similarly, we shall say
$v' \in V'$ is a weak-pivot $p$ for $p'$ if
\begin{equation}
d(p', H) >  d(v', H).
\end{equation}

In an iteration of Triangle Algorithm II a given pair $(p_k,p'_k) \in K \times K'$ may or many not be a witness pair. The algorithm  searches for a weak-pivot or a pivot in order to reduce the current gap $\delta_k=d(p_k,p'_k)$ until $\epsilon$-approximate solutions to both the distance and supporting hyperplanes problems are reached.

The correctness of the Triangle Algorithm relies on the following distance duality.

\begin{thm}  \label{thm1} {\rm (Distance Duality~\cite{trianglesvm})}  $K \cap K' \not = \emptyset$ if and only if for each $(p, p')  \in K \times K'$, either
there exists  $v \in S$ such that $d(p, v) \geq d (p', v)$, or there exists $v' \in S'$ such that $d(p', v') \geq d(p, v')$. $\Box$
\end{thm}

Suppose $d(K,K')=d(p_*, p'_*)$, where $(p_*, p'_*) \in K \times K$.  Then  if $H_{p_*}$  and $H_{p'_*}$  are orthogonal hyperplanes  to the line segment $p_*p'_*$ at $p_*$ and $p'_*$ respectively, they are optimal supporting hyperplanes to $K$ and $K'$, respectively. In other words,  $d(K,K')=d(p_*, p'_*)=d(H_{p_*}, H_{p'_*})$.

Table \ref{table:complexity} summarizes the complexity of Triangle Algorithms I and II in solving the optimal hyperplane problem.(See \cite{trianglesvm})
\begin{table}[H]
	\renewcommand{\arraystretch}{1.0}
	\centering
\scalebox{0.92
}{
\begin{tabular}{|l|l|l|c|}

\hline
Complexity  of  computing &  ~~~~~~~Intersection & ~~~~~~~Separation & Distance and Support
\\
 $\epsilon$-approximation solution& $~~~~~~K \cap K' \not = \emptyset $ & $~~~~~~K \cap K'= \emptyset $ &  $\delta_*=d(K,K')$
\\
	\hline

$K = conv(\{v_1, \dots, v_n\})$  &  & &
\\
& $~~~~~~~O\big ( mn\frac{1}{\epsilon^2} \big)$ & $~~~~~~O \big  (mn \big (\frac{\rho_*}{\delta_*} \big)^2\big)$ & $~O \big (mn \big (\frac{\rho_*} {\delta_*\epsilon} \big)^2 \big )$
\\
complexity w. preprocessing  & $~~~~O\big ((m+n)\frac{1}{\epsilon^2} \big)$ & $~~O \big  ((m+n) \big (\frac{\rho_*}{\delta_*} \big)^2\big)$ & $~O \big ((m+n) \big (\frac{\rho_*} {\delta_*\epsilon} \big)^2 \big )$
\\
$K' = \{p'\}$  &  & &
\\
\hline
$K = conv(\{v_1, \dots, v_n\})$ & & &
\\

$N = \max \{n,n'\}$& $~~~~~~~O\big (m N \frac{1}{\epsilon^2} \big)$ & $~~~~~O \big (m N\big (\frac{\rho_*}{\delta_*} \big )^2 \big )$ & $O \big ( m N \big (\frac{\rho_*} {\delta_*\epsilon} \big)^2 \ln \frac {\rho_*}{\delta_*}  \big )$ 	
\\
complexity w. preprocessing & $~~~~O\big ((m +N )\frac{1}{\epsilon^2} \big)$ & $~~O \big ((m + N)\big (\frac{\rho_*}{\delta_*} \big )^2 \big )$ & $O \big ( (m + N) \big (\frac{\rho_*} {\delta_*\epsilon} \big)^2 \ln \frac {\rho_*}{\delta_*}  \big )$ 	
\\
$K' = conv(\{v'_1, \dots, v'_{n'}\})$ & & &
\\
\hline

	\end{tabular}
}
\caption{The complexities of Triangle Algorithms I and II. $\rho_*$ is maximum of diameters of $K$ and $K'$.}
\label{table:complexity}
\end{table}

\subsection{Formal definition of Triangle Algorithm I}
In this section we describe the details of {\it Triangle Algorithm I}. This is a generalization of the original Triangle Algorithm in the case when $V = \{v_1,\dots,v_n\}$ and $V'= \{v_1',\dots,v_{n'}'\}$.

The algorithm searches for a triangle $\triangle pp'v'$ where $v' \in V'$, such that $d(p',v') \geq d(p,v')$; or
a triangle $\triangle pp'v$ where $v \in V$, such that $d(p,v) \geq d(p',v)$.  Given that such triangle exists, it uses $v$ or $v'$ as a pivot to bring $p,p'$ in current iterate $(p,p') = (p_k,p_k') \in K \times K'$  closer to each other by generating either a new iterate $p_{k+1} \in K$, or new iterate $p'_{k+1} \in K'$ such that if we denote the new iterate by $(p_{k+1}, p'_{k+1})$,
$d(p_{k+1}, p'_{k+1}) < d(p_k,p'_k)$.

Given three points $x,y,z \in \mathbb{R}^m$ such that $d(y,z) \geq d(x,z)$. Let $nearest(x; yz)$ be the nearest point to $x$ on the line segment joining  $y$ to $z$.

We have
Given three points $x,y,z \in \mathbb{R}^m$, let the {\it step-size} be
\begin{equation}
\alpha = \frac{(x-y)^T(z-y)}{d^2(y,z)}.
\end{equation}
Then
\begin{equation} \label{pdp}
nearest(x; yz)=
\begin{cases}
(1-\alpha)y + \alpha z, &\text{if $\alpha \in [0,1]$;}\\
z, &\text{otherwise.}
\end{cases}
\end{equation}

Here we describe Triangle Algorithm I for testing if two finite convex hulls $K, K'$ intersect. It computes a pair $(p,p')  \in K \times K'$ such that either $d(p,p')$ is to within a prescribed tolerance, or it is a witness pair. It assumes we are given points $(p_0, p_0') \in K \times K'$ and $\epsilon \in (0,1)$.

\begin{center}
\begin{tikzpicture}
%\begin{center}
\node [mybox] (box){%
    \begin{minipage}{0.9\textwidth}
{\bf  Triangle Algorithm I ($(p_0, p'_0) \in K \times K'$, $\epsilon \in (0,1)$)}\

\begin{itemize}
\item
{\bf Step 0.} Set $p=v=p_0$, $p'=v'= p_0'$.

\item
{\bf Step 1.} If $d(p,p') \leq \epsilon d(p,v)$, or $d(p,p') \leq \epsilon d(p',v')$, stop.

\item
{\bf Step 2.}  Test if there exists $v \in V$ that is a $p$-pivot for $p'$, i.e.
\begin{equation} \label{pivotsx}
2v^T(p'-p) \geq \Vert p' \Vert^2 -  \Vert p \Vert^2
\end{equation}
If such pivot exists, set $p \leftarrow nearest(p'; pv)$, and go to Step 1.

\item
{\bf Step 3.}  Test if there exists $v' \in V'$ that is a $p'$-pivot for $p$, i.e.
\begin{equation} \label{pivotsxx}
2v'^T(p-p') \geq \Vert p \Vert^2 -  \Vert p' \Vert^2
\end{equation}
If such pivot exists, set $p' \leftarrow nearest(p; p'v')$, and go to Step 1.

\item

{\bf Step 4.} Output $(p,p')$ as a witness pair, stop ($K \cap K' = \emptyset$).

\end{itemize}
\end{minipage}};
\end{tikzpicture}
\end{center}

\subsection{Formal definition of Triangle Algorithm II}
Triangle Algorithm II begins with a witness pair $(p, p') \in K \times K'$, then it computes an $\epsilon$-approximate solution to the distance problem.  Since $(p, p')$ is a witness pair there exists no $p'$-pivot for $p$, or a $p$-pivot for $p'$. However, if  $(p, p')$ is not
already an $\epsilon$-approximate solution to $\delta_*$, the algorithm makes use of a {\it weak-pivot}.

Given a  pair $(p,p') \in K \times K'$, the orthogonal bisector hyperplane of the line segment $pp'$ is
\begin{equation}
H=\{x \in \mathbb{R} ^m:  \quad h^Tx = a \},
\quad h=p-p', \quad a = \frac{1}{2} (p^Tp-p'^Tp').
\end{equation}
If $(p,p')$ is a witness pair then
\begin{equation}
K \subset H_{+}=\{x \in \mathbb{R} ^m:  h^Tx >  a  \}, \quad K' \subset   H_{-}=\{x \in \mathbb{R}^m:  h^Tx < a \}. ~~~\Box
\end{equation}

Let
\begin{equation} \label{vv'}
v={\rm argmin}\{h^Tx: x \in V\},  \quad v'={\rm argmax}\{h^Tx: x \in V'\}.
\end{equation}

\begin{equation} \label{hh}
H_v= \{x: h^Tx= h^Tv\}, \quad H_{v'}= \{x: h^Tx= h^Tv'\}.
\end{equation}

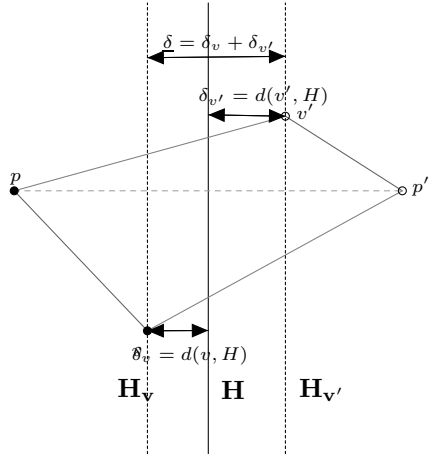
\begin{figure}[H]
\centering
\definecolor{yqyqyq}{rgb}{0.5019607843137255,0.5019607843137255,0.5019607843137255}
\definecolor{aqaqaq}{rgb}{0.6274509803921569,0.6274509803921569,0.6274509803921569}
\definecolor{wqwqwq}{rgb}{0.3764705882352941,0.3764705882352941,0.3764705882352941}
\begin{tikzpicture}[line cap=round,line join=round,>=triangle 45,x=1.0cm,y=1.0cm,xscale=1,yscale=1]
\clip(2.,1.) rectangle (8.5,7.);
\draw (5.16,1.) -- (5.16,7.);
\draw [color=wqwqwq] (2.58,4.5)-- (4.350943588315216,2.637422144585118);
\draw [color=wqwqwq] (7.74,4.5)-- (6.185153241396304,5.492683097596657);
\draw (5.209241696960757,2.099384562123556) node[anchor=north west] {$\mathbf{H}$};
\draw [dash pattern=on 2pt off 2pt,color=aqaqaq] (2.58,4.5)-- (7.74,4.5);
\draw [line width=0.4pt,color=yqyqyq] (2.58,4.5)-- (6.185153241396304,5.492683097596657);
\draw [line width=0.4pt,color=yqyqyq] (4.350943588315216,2.637422144585118)-- (7.74,4.5);
\draw [dash pattern=on 1pt off 1pt] (6.185153241396304,1.) -- (6.185153241396304,7.);
\draw [dash pattern=on 1pt off 1pt] (4.350943588315216,1.) -- (4.350943588315216,7.);
\draw (6.242870946333856,2.125887876210046) node[anchor=north west] {$\mathbf{H_{v'}}$};
\draw (3.831069364463292,2.125887876210046) node[anchor=north west] {$\mathbf{H_v}$};
\draw [->,line width=0.4pt] (4.350943588315216,2.637422144585118) -- (5.16,2.6415991766744984);
\draw [->] (5.16,2.6415991766744984) -- (4.350943588315216,2.637422144585118);
\draw [->,line width=0.4pt] (6.185153241396304,5.492683097596657) -- (5.16,5.510785217729497);
\draw [->,line width=0.4pt] (5.16,5.510785217729497) -- (6.185153241396304,5.492683097596657);
\draw [->,line width=0.4pt] (4.350943588315216,6.27107426330874) -- (6.185153241396304,6.289176383441578);
\draw [->,line width=0.4pt] (6.185153241396304,6.289176383441578) -- (4.350943588315216,6.27107426330874);
\begin{scriptsize}
\draw [fill=black] (2.58,4.5) circle (1.5pt);
\draw[color=black] (2.5986652594415203,4.656954371469817) node {$p$};
\draw [color=black] (7.74,4.5) circle (1.5pt);
\draw[color=black] (7.992089676042177,4.564192772167103) node {$p'$};
\draw [color=black] (6.185153241396304,5.492683097596657) circle (1.5pt);
\draw[color=black] (6.454897459025774,5.558067050410469) node {$v'$};
\draw [fill=black] (4.350943588315216,2.637422144585118) circle (1.5pt);
\draw[color=black] (4.215367418717393,2.3511660459452086) node {$v$};
\draw[color=black] (4.9177052420093705,2.3114110748154744) node {$\delta_v = d(v,H)$};
\draw[color=black] (5.89832786320949,5.770093563102387) node {$\delta_{v'} = d(v',H)$};
\draw[color=black] (5.309241696960757,6.459179729351121) node {$\underline\delta =\delta_v+\delta_{v'}$};
\end{scriptsize}
\end{tikzpicture}
\caption{Distances from the separating hyperplane}
\label{fig:ta2def}
\end{figure}

Then the hyperplanes $H_v$ and $H_{v'}$ give supporting hyperplanes to $K$ and $K'$, respectively, and the distance between them, $d(H_v, H_{v'})$ is a lower bound to $\delta_*$.
Specifically, let
\begin{equation} \label{deltavv}
\delta_v=d(v, H), \quad  \delta_{v'}= d(v', H).
\end{equation}
Then if
\begin{equation} \label{deltalow}
\underline \delta= \delta_v + \delta_{v'},
\end{equation}
we have
\begin{equation} \label{maineq}
d(H_v, H_{v'})=\underline \delta= \frac{h^Tv-h^Tv'}{\Vert h \Vert},
\end{equation}
and
\begin{equation} \label{maineq2}
\underline \delta \leq \delta_* \leq \delta=d(p,p').
\end{equation}
Figure \ref{fig:ta2def}, gives a geometric interpretation of the distances.
Let  $(p,p')$ be a witness pair and consider $v, v'$  as defined in (\ref{vv'}). Also let $H_v$, $H_{v'}$ and  $H$ be the hyperplanes defined earlier. Let $\delta=d(p,p')$,
$\rho=d(p,v)$, and $\rho'=d(p',v')$. Let
$\underline \delta=d(H_v, H_{v'})$,  and $E=\delta - \underline \delta$. We shall say $(p,p')$ gives a {\it strong}  $\epsilon$-{\it approximate solution} to $\delta_*$ if either
\begin{equation}
E \leq \epsilon \rho, \quad {\rm or} \quad E \leq \epsilon \rho'.
\end{equation}

Given a witness pair $(p,p')$, let $\delta = d(p,p')$ and $\delta_v$, $\delta_{v'}$ be as defined in (\ref{deltavv}).  Define
\begin{equation}  \label{Ev}
E_v= (\frac{1}{2} \delta - \delta_v), \quad E_{v'}= (\frac{1}{2}  \delta - \delta_{v'}).
\end{equation}
Clearly,
\begin{equation}
E= \delta - \underline \delta = E_{v} + E_{v'}.
\end{equation}

The input to \emph{Triangle Algorithm II} is a witness pair $(p,p')$. It computes a new witness pair $(p,p')$ that gives an $\epsilon$-approximate solution to $\delta_*$, as well as a pair $(v,v') \in K \times K'$, where the hyperplanes  parallel to the orthogonal bisecting hyperplane of $pp'$ passing through $v,v'$ form a pair of  supporting hyperplanes, giving  an $\epsilon$-approximate solution to the supporting hyperplanes problem.(See box)

\begin{center}
\begin{tikzpicture}
%\begin{center}
\node [mybox] (box){%
    \begin{minipage}{0.9\textwidth}
{\bf  Triangle Algorithm II ($(p,p') \in K \times K'$, a witness pair, $\epsilon \in (0,1)$)}\

\begin{itemize}
\item
{\bf Step 1.} Set $h=p-p'$. Compute
$$v={\rm argmin}\{h^Tx: x \in V\}, \quad v'={\rm argmax}\{h^Tx: x \in V'\}.$$
$$ E= \delta- \underline \delta, \quad  \delta=d(p,p'), \quad \underline \delta= ({h^Tv-h^Tv'})/{\Vert h \Vert}.$$

\item
{\bf Step 2.} If $E \leq  \epsilon \rho$,  or
$E\leq  \epsilon \rho'$, with $\rho=d(p,v)$, $\rho'=d(p',v')$, output $(p,p')$, $(H_v,H_{v'})$, stop.

\item
{\bf Step 3.}  If $E_v > \frac{1}{2} \epsilon \rho $, compute $p \leftarrow nearest(p'; pv)$, go to Step 5.

\item
{\bf Step 4.}  If $E_{v'} > \frac{1}{2}\epsilon \rho'$,  compute  $p' \leftarrow nearest(p; p'v')$, go to Step 5.

\item

{\bf Step 5.} Call Triangle Algorithm I with $(p, p')$ as input. Go to Step 1.

\end{itemize}

    \end{minipage}};
%\end{center}
\end{tikzpicture}
\end{center}

In section \ref{section:implementation} we discuss certain notes which are useful in an efficient implementation of the Triangle Algorithm.

\section{Lagrange Duality and SMO algorithm}\label{smosection}
The SMO algorithm relies on the fact that the optimal hyperplane $w$ can be expressed as a convex combination of the input vectors i.e. $w = \sum_{i=1}^{m} {\alpha_i} y^{(i)} x^{(i)}$. It then performs a coordinate descent over the parameters $\alpha_1 \dots \alpha_m$ to find the optimal hyperplane.
The remainder of this section describes the Lagrange duality for the optimization problem and SMO algorithm.

\subsection{Lagrange Duality}
Given the training set $\{x^{(i)}, \quad i = 1,\dots,m\}$ and their convex hull membership denoted by $y_i$, we define the {\it functional margin} $\gamma$ to be
\begin{equation} \label{eq:6}
    \gamma = y_i (w^Tx^{(i)} + b) \quad y_i \in \{-1,1\}
\end{equation}

To keep the functional margin to be large we need $w^Tx^{(i)} + b$ to be large positive number in case of $y^{(i)}=1$ and a large negative number in case of $y^{(i)}=-1$.
However, choosing $(2w,2b)$ instead of $(w,b)$ would also increase the functional margin without doing anything meaningful.

Enforcing $||w||=1$ ensures that the functional margin equals the geometric margin, and guarantees that all geometric margins are at least equal to $\gamma$. 

To find the maximum of the minimum margins, equation(\ref{eq:6}) can be written as
\begin{equation}
\begin{split}
    & \underset{\gamma, w, b}{max} ~\gamma \\
    s.t.\quad & y_i (w^Tx^{(i)} + b) \ge \gamma, \quad i=1,\dots,m \\
    & \Vert w \Vert = 1
\end{split}
\end{equation}

As $\Vert w \Vert = 1$ leads to a non-convex constraint, we formulate the following problem,
\begin{equation}
\begin{split}
    & max_{ \gamma, w, b} \frac{\gamma}{||w||} \\
    s.t.\quad & y^{(i)} (w^Tx^{(i)} + b) \ge \gamma, \quad i=1,\dots,m \\
\end{split}
\end{equation}

Without l.o.g. we allow $\gamma = 1$. Maximizing $\frac{\gamma}{||w||}$ or $\frac{1}{||w||}$ is equivalent to minimizing $||w||^2$. 
We now have an optimization problem that can be efficiently solved and the solution to which gives an {\it optimal-margin classifier}.

\begin{equation} \label{eq:10}
\begin{split}
    & \underset{w,b}{min} ~ \frac{1}{2} {\Vert w \Vert}^2 \\
    &s.t.\   y_i (w^Tx^{(i)} + b) \ge 1, \quad i=1,\dots,m
\end{split}
\end{equation}
In the machine learning literature, when their convex hulls are disjoint, the hard margin SVM problem is to compute a pair of supporting hyperplanes with maximum margin. This optimization is infeasible when the convex hulls $K$ and $K'$ intersect, i.e., $K \cap K' \neq \emptyset$. In such cases a soft margin SVM formulates the problem as a convex programming whose optimal solution provides a pair of supporting hyperplanes, necessarily allowing some mis-classifications.

In that case we define a relaxation coefficient for the separation and penalize the objective function with respect to the relaxation coefficient resulting in the soft margin formulation
\begin{equation}
\begin{split}
& \underset{w,b}{min} \frac{1}{2} {\Vert w \Vert}^2 + C\sum_{i=1}^m \xi_i\\
&s.t.\   y_i (w^Tx^{(i)} + b) \ge 1 - \xi_i, i=1,\dots,m \\
& \xi_i \ge 0, i=1,\dots,m
\end{split}
\end{equation}
The parameter C controls the weighting between the twin goals of minimizing $||w||^2$ and ensuring the most examples have a functional margin at least 1.

The Lagrangian for the optimization problem, as described in ~\cite{vapnik} turns out to be 

\begin{equation}\label{ldual}
    L(w,b,\xi,\alpha,r)=\frac{1}{2}w^Tw  + C\sum_{i=1}^m \xi_i -\sum_{i=1}^m \alpha_i[y_i(x^Tw+b)-1+\xi_i]-\sum_{i=1}^m r_i\xi_i
\end{equation}
Taking the derivative of the Lagrangian and setting it yo zero we get,
\begin{equation} \label{lderiv}
\begin{split}
    w = \sum_{i=1}^{m} {\alpha_i} y_i x^{(i)} \\
    \sum_{i=1}^m \alpha_i{y_i} = 0 
\end{split}
\end{equation}
Plugging (\ref{lderiv}) back to (\ref{ldual}) gives way to the Wolfe dual of the problem
\begin{equation}\label{wdual}
\begin{split}
    & \underset{\alpha}{max} W(\alpha) = \sum_{i=1}^m {\alpha_i} -\frac{1}{2} \sum_{i,j=1}^m y_i y_j{\alpha_i}{\alpha_j}{x^{(i)}}{x^{(j)}}\\
&s.t.\   0 \le \alpha_i \le C, i=1,\dots,m \\
& \sum_{i=1}^m \alpha_i{y_i} = 0
\end{split}
\end{equation}
The KKT conditions for convergence are
\begin{equation}\label{kkt}
\begin{split}
    \alpha_i = 0 &\implies y_i(wx^{(i)}+b) > 1 \\
    \alpha_i = C &\implies y_i(wx^{(i)}+b) < 1 \\
    0 < \alpha_i < C &\implies  y_i(wx^{(i)}+b) = 1
\end{split}
\end{equation}
Notice that (\ref{wdual}) does not solve for the constant $b$ explicitly. Once we have the optimal $w^*$, $b^*$ can be calculated as

\begin{equation}
    b^* = - \frac{1}{2}(~\underset{i:y_i=-1}{max} ~ w^{*T}x^{(i)} + ~\underset{i:y_i=1}{min} ~ w^{*T}x^{(i)})
\end{equation}

\subsection{The SMO algorithm}

The SMO algorithm performs a coordinate descent over the set of all $\alpha_{i}s$. The constraint in (\ref{lderiv}) forces the sum of the products $\alpha_i{y_i}$ to be zero. For
\begin{equation}
    \alpha_1  = \frac{1}{{y_1}} (-\sum_{i=2}^m \alpha_i{y_i})
\end{equation}
coordinate descent cannot be performed by adjusting a single value of $\alpha_i$. We can perform coordinate descent between a pair $(\alpha_i,\alpha_j)$ such that their sum is constant. 

The SMO algorithm can be described as following:
\begin{itemize}
    \item Select a pair $(\alpha_i,\alpha_j)$ to update.
    \item Reoptimize $W(\alpha)$ with respect to $(\alpha_i,\alpha_j)$, while holding other $\alpha_k$ ($k \ne i,j$) fixed.
\end{itemize}

Specifically we choose $\alpha_i$ and $\alpha_j$ such that
\begin{equation}
\begin{split}
    \alpha_i{y_i} + \alpha_j{y_j} &= -\sum_{k=1}^{i-1} \alpha_k{y_k}-\sum_{k=i+1}^{j-1} \alpha_k{y_k}-\sum_{k=j+1}^{m} \alpha_k{y_k} \\
    \alpha_i{y_i} + \alpha_j{y_j} &= \varsigma
\end{split}
\end{equation}
where $\varsigma$ is some constant. This forces a line constraint. Moreover the KKT conditions(\ref{kkt}) enforces the following box constraint.

\begin{figure}[H] 
\centering
\begin{tikzpicture}
\draw[-> ] (-0.5,0) -- (3.5,0);
\draw[->] (0,-0.5) -- (0,3.5);
\draw[gray, thin] (-0.5,3) -- (3.5,3);
\draw[densely dotted] (-0.5,2) -- (3.5,2);
\draw[gray, thin] (3,-0.5) -- (3,3.5);
\draw[gray, thin] (0.5,-0.5) -- (3.5,2.5);
\filldraw[black] (3,2) circle (1pt) node[anchor=west] {$\alpha_1{y^{(1)}} + \alpha_2{y^{(2)}} = \varsigma$};
\filldraw[black] (1.5,0) circle (0pt) node[anchor=north] {$\alpha_1$}; 
\filldraw[black] (0,1.5) circle (0pt) node[anchor=east] {$\alpha_2$}; 
\filldraw[black] (0,2) circle (1pt) node[anchor=south east] {H}; 
\filldraw[black] (0,3) circle (1pt) node[anchor=south east] {C}; 
\filldraw[black] (3,0) circle (1pt) node[anchor=north west] {C}; 
\filldraw[black] (0,0) circle (1pt) node[anchor=north east] {L}; 
\end{tikzpicture}
\caption{Constraints for $(\alpha_i,\alpha_j)$ and $y^{(i)}= y^{(j)}$}
\end{figure}
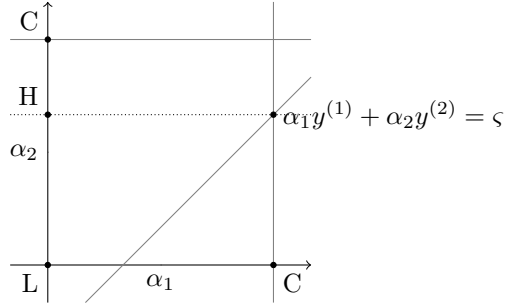

Given the choice of two $\alpha_i$s to optimize we first compute the lower and upper bounds as following:
\begin{equation}
    \bullet \qquad y_i\neq y_j, \qquad L=max(0,\alpha_j-\alpha_i), H=min(C,C+\alpha_j-\alpha_i)
\end{equation}
\begin{equation}
    \bullet \qquad y_i= y_j, \qquad L=max(0,\alpha_j+\alpha_i-C), H=min(C,\alpha_j+\alpha_i)
\end{equation}

Now we want the optimal value for $\alpha_j$ to maximize the objective function. If this value ends up lying
outside the bounds L and H, we simply clip the value of $\alpha_j$ to lie within this range. The update for $\alpha_j$ as discussed in ~\cite{plattsmo} is
\begin{equation}
    \alpha_j = \alpha_j + \frac{y_j(E_i-E_j)}{\eta}
\end{equation}
where
\begin{equation}
    f(x) = \sum_{i=1}^m \alpha_i y_i (x^{(i)}.x) + b
\end{equation}
\begin{equation}
    E_k = f(x^{(k)})-y_k
\end{equation}
\begin{equation}
    \eta = 2x^{(i)}x^{(j)} - x^{(i)}x^{(i)}-x^{(j)}x^{(j)}
\end{equation}
If the optimal value of $\alpha_j$ is not within the bounds we clip it as following
\[
\alpha_j = 
\begin{cases}
    H, & \text{if } \alpha_j > H \\
    \alpha_j, & \text{if } L \le \alpha_j \le H \\
    L, & \text{if } \alpha_j < L
\end{cases}
\]

After solving for $\alpha_j$ we solve for $\alpha_i$ defined as
\begin{equation}
    \alpha_i = \alpha_i + y_i y_j(\alpha_j^{(old)}-\alpha_j)
\end{equation}
Where $\alpha_j^{(old)}$ is the value of $\alpha_j$ from the previous iteration.

Next we select the threshold b to satisfy the KKT conditions. If $\alpha_i$ is not at bounds (i.e. $0 < \alpha_i < C$) then the following threshold $b_1$ is valid
\begin{equation}
    b_1 = b - E_i - y_i(\alpha_i - \alpha_{i}^{(old)})(x^{(i)}x^{(i)})  - y_j(\alpha_j - \alpha_{j}^{(old)})(x^{(i)}x^{(j)})
\end{equation}

Similarly if $\alpha_j$ is not at bounds the following threshold $b_2$ is valid
\begin{equation}
    b_1 = b - E_j - y_i(\alpha_i - \alpha_{i}^{(old)})(x^{(i)}x^{(j)})  - y_j(\alpha_j - \alpha_{j}^{(old)})(x^{(j)}x^{(j)})
\end{equation}

If both the thresholds are not at bounds, then all thresholds between $b_1$ and $b_2$ satisfy the KKT conditions. This gives the following update equation

\[
b = 
\begin{cases}
    b_1 & \text{if } 0 < \alpha_i < C \\
    b_2 & \text{if } 0 < \alpha_j < C \\
    (b_1 + b_2)/2 & \text{otherwise}
\end{cases}
\]

Each iteration of the SMO algorithm has two loops. The outer loops picks $\alpha_1$ that corresponds to a sample violating the KKT conditions. The inner loop then picks another $\alpha_2$ such that the error $E_2 - E_1$ is maximum.

\begin{center}
\begin{tikzpicture}
%\begin{center}
\node [mybox] (box){%
    \begin{minipage}{0.9\textwidth}
{\bf  SMO algorithm (Main loop) }\

\begin{itemize}
\item
{\bf Step 1.} Initialize $\alpha_i = 0$, $\forall$ i, b=0, passes=0, numChanged=0, examineAll=0

\item
{\bf Step 2.} while $(numChanged > 0$ $\|$ $examineAll)$
\begin{itemize}
    \item[$\circ$] numChanged=0
    \item[$\circ$] if (examineAll)
    \begin{itemize}
        \item[$\circ$] loop I over all training examples
        \begin{itemize}
            \item[$\circ$] numChanged=numChanged + examineExample(I)
        \end{itemize}
    \end{itemize}
    \item[$\circ$] else
    \begin{itemize}
        \item[$\circ$] loop I over all training examples where alpha is not 0 or not C
        \begin{itemize}
            \item[$\circ$] numChanged=numChanged + examineExample(I)
        \end{itemize}
    \end{itemize}
     \item[$\circ$] if (examineAll == 1)
    \begin{itemize}
        \item[$\circ$] examineAll = 0
    \end{itemize}
    \item[$\circ$] if (examineAll == 0)
    \begin{itemize}
        \item[$\circ$] examineAll = 1
    \end{itemize}
\end{itemize}

\end{itemize}

    \end{minipage}};
%\end{center}
\end{tikzpicture}
\end{center}

\begin{center}
\begin{tikzpicture}
%\begin{center}
\node [mybox] (box){%
    \begin{minipage}{0.9\textwidth}
{\bf  examineExample(i2) }\
\begin{itemize}
\item
{\bf Step 1.} y2= target(i2), alph2=Lagrange multiplier for i2, E2 = error(i2) - y2, r2=E2*y2
\item
{\bf Step 2.} if $(r2 < -tol$ \& $alph2 < C)$  $\|$ $(r2 > tol$ \& $alph2 > 0)$
\begin{itemize}
    \item[$\circ$] if (number of non-zero and non-C alpha $>$ 1) 
    \begin{itemize}
        \item[$\circ$] i1 = result of second choice heuristics
        \item[$\circ$] if takestep(i1,i2) return 1
    \end{itemize}
    \item[$\circ$] loop over all non-zero and non-C alpha 
    \begin{itemize}
        \item[$\circ$] i1 = identity of current alpha
        \item[$\circ$] if takestep(i1,i2) return 1
    \end{itemize}
    \item[$\circ$] loop over all possible i1 
    \begin{itemize}
        \item[$\circ$] i1 = loop variable
        \item[$\circ$] if takestep(i1,i2) return 1
    \end{itemize}
    \item[$\circ$] return 0
\end{itemize}

\end{itemize}
\end{minipage}};
%\end{center}
\end{tikzpicture}
\end{center}

\begin{center}
\begin{tikzpicture}
%\begin{center}
\node [mybox] (box){%
    \begin{minipage}{0.9\textwidth}
{\bf  takeStep(i1,i2) }\
\begin{itemize}
\item
{\bf Step 1.} Clip and update $alph1$ and $alph2$
\item
{\bf Step 2.} Update threshold to reflect change in Lagrange multiplier

\item
{\bf Step 3.} Update weight vector to reflect change in alph1 and alph2
\item
{\bf Step 4.} Update error cache in using Lagrange multiplier

\end{itemize}
\end{minipage}};
%\end{center}
\end{tikzpicture}
\end{center}

The SMO algorithm however, does not answer the question of separability of two convex sets. A hard margin classification can be enforced by setting the value of C to infinity, but in case of overlapping convex hulls it doesn't converge.

\section{Testing intersection or separation of Convex Hulls}\label{section:intersection}

Testing for intersection of two finite Convex Hulls is a general case for testing if a point lies inside the Convex Hull. The Triangle Algorithm is efficient in answering the convex hull problem.~\cite{triangleperformance}

In case of linearly separable sets we report the number of iterations it takes for the Triangle Algorithm to converge. The value of $\epsilon = 0.001$ was used and the maximum number of iterations was set to be $10^4$.

In the experimental setup we generated points in $V=\{v_1 \dots v_n\}$ and $V'=\{v_1' \dots v_n'\}$ from two unit balls. Let $K=conv(V)$ and $K'=conv(V')$. We translated the smaller ball along a random direction by $\frac{9}{10}max\{diam(K),diam(K')\}$ units.
Where
\begin{equation} \label{eq:12}
    diam(K) = max\{d(v_i,v_j): \quad (v_i,v_j) \in K\}
\end{equation}
and $d(v_i,v_j)$ is the euclidean distance between $v_I$ and$v_j$.
The number of points in each of the set K and K' were 5000. Figure \ref{fig:triangleIntersection} shows the performance in the number of iterations and time.

\begin{table}[H]
\centering
\pgfkeys{/pgf/number format/.cd, precision=3}
\pgfplotstabletypeset[
    %columns={dimension,triteration,trtime,trsparsity,trdist,smoiteration,smotime,smosparsity,smodist},
    columns={dimension,triteration,trtime},
    every head row/.style={%
        before row={\toprule%
            &\multicolumn{2}{c}{Triangle}  \\
            },
        after row=\hline
    },
    columns/dimension/.style={column name=Dimensions, column type={|l}},
    columns/triteration/.style={column name=Iterations, column type={|l}},
    columns/trtime/.style={column name=Time(sec), column type={|l|}},
    every last row/.style={after row=\hline},
]\triangleSeparation
\caption{Triangle Algorithm performance for set intersection}
\end{table}
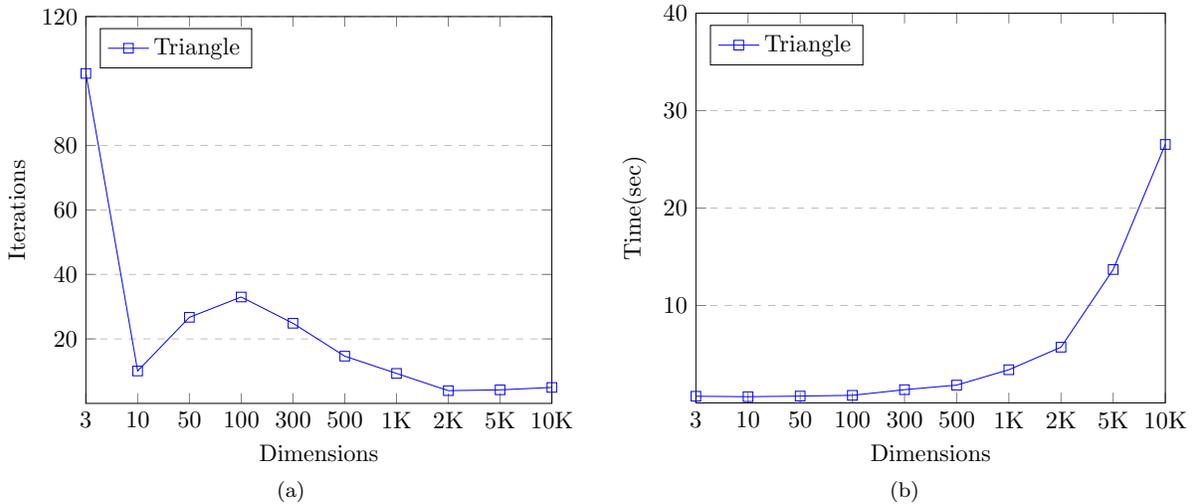
\begin{figure}
\subfloat[]{{
\resizebox {0.48\columnwidth} {!} {
\begin{tikzpicture}
\begin{axis}[
    title={},
    xlabel={Dimensions},
    ylabel={Iterations},
    xmin=1, xmax=10,
    ymin=0, ymax=120,
    xtick= {1,2,3,4,5,6,7,8,9,10},
    xticklabels from table ={\dimensionlabel}{dimension},
    xticklabel style={/pgf/number format/fixed,
                  /pgf/number format/precision=3},
    ytick={20,40,60,80,120},
    legend pos=north west,
    ymajorgrids=true,
    grid style=dashed,
    enlargelimits=false
]
 
\addplot[color=blue,mark=square] table[x=xlabel,y=triteration] {\triangleSeparation};
\addlegendentry{Triangle}
\end{axis}
\end{tikzpicture}
}
}}
\subfloat[]{{
\resizebox {0.48\columnwidth} {!} {
\begin{tikzpicture}
\begin{axis}[
    title={},
    xlabel={Dimensions},
    ylabel={Time(sec)},
    xmin=1, xmax=10,
    ymin=0, ymax=40,
    xtick= {1,2,3,4,5,6,7,8,9,10},
    xticklabels from table ={\dimensionlabel}{dimension},
    xticklabel style={/pgf/number format/fixed,
                  /pgf/number format/precision=3},
    ytick={10,20,30,40},
    legend pos=north west,
    ymajorgrids=true,
    grid style=dashed,
    enlargelimits=false
]
 
\addplot[color=blue,mark=square] table[x=xlabel,y=trtime] {\triangleSeparation};
\addlegendentry{Triangle}
\end{axis}
\end{tikzpicture}
}
}}
\caption{Performance of Triangle Algorithm for testing intersection} \label{fig:triangleIntersection}
\end{figure}

\section{Testing approximation of Distance and Optimal support}\label{section:optimal}
We compared the performance of Triangle Algorithm with SMO for calculating the optimal support, given that the convex hulls are separate.
To enforce a hard-margin classification we set the constant $C=Infinity$ for SMO. The value of $\epsilon = 0.001$ was used for both algorithms. Additionally the maximum number of iterations were set to be $10^4$ for both algorithms.

\subsection{Comparison Based on Dimension}
In the experimental setup we generated points in $V=\{v_1 \dots v_n\}$ and $V'=\{v_1' \dots v_{n'}'\}$ from two unit balls. Let $K=conv(V)$ and $K'=conv(V')$. We translated the smaller ball along a random direction by $max\{\frac{11}{10}diam(K),\frac{11}{10}diam(K')\}$ units. Where diam is defined in (\ref{eq:12}). The lower bound for the distance between the two convex hulls, $K=conv(V)$ and $K'=conv(V')$, becomes  $max\{\frac{1}{10}diam(K),\frac{1}{10}diam(K')\}$. 
The number of points in each of the set $V$ and $V'$ were 5000.
In our results we also report the sparsity and mean estimated distance calculated by both SMO and Triangle Algorithm.
From (\ref{lderiv}) we know that the hyperplane $w$ is a sparse combination of the points in $V$ and $V'$. The table shows the number of points it took to represent the optimal $w$.
The distance is the reported distance between the sets $V$ and $V'$, which is calculated as
\begin{equation}
    d = \frac{1}{\Vert w \Vert}(-max\{w^Tv + b: \quad v \in V\} + min\{w^Tv' + b: \quad v' \in V'\})
\end{equation}
Figure \ref{fig:opsupport} shows the performance of both algorithms in terms of iterations and time.
%\begin{center}

\begin{table}
\centering
\resizebox{\columnwidth}{!}{
\pgfkeys{/pgf/number format/.cd, precision=2}
\pgfplotstabletypeset[
    columns={dimension,triteration,trtime,trsparsity,trdist,smoiteration,smotime,smosparsity,smodist},
    every head row/.style={%
        before row={\toprule%
            &\multicolumn{4}{c}{Triangle}  
            &\multicolumn{4}{c}{SMO} \\
            },
        after row=\hline
    },
    columns/dimension/.style={column name=Dimensions, column type={|l}},
    columns/triteration/.style={column name=Iterations, column type={|l}},
    columns/trtime/.style={column name=Time(sec), column type={|l}},
    columns/trsparsity/.style={column name=Sparsity, column type={|l}},
    columns/trdist/.style={column name=Distance, column type={|l}},
    columns/smoiteration/.style={column name=Iterations, column type={|l}},
    columns/smotime/.style={column name=Time(sec), column type={|l}},
    columns/smosparsity/.style={column name=Sparsity, column type={|l}},
    columns/smodist/.style={column name=Distance, column type={|l|}},
    every last row/.style={after row=\hline},
]\triangleDimensionData
}
\caption{Performance based on dimension}
\end{table}
\begin{figure}
\subfloat[]{{
\resizebox {0.48\columnwidth} {!} {
\begin{tikzpicture}
\begin{axis}[
    title={Comparison of iterations triangle/SMO},
    xlabel={Dimensions},
    ylabel={Iterations},
    xmin=1, xmax=10,
    ymin=0, ymax=800,
    xtick= {1,2,3,4,5,6,7,8,9,10},
    xticklabels from table ={\dimensionlabel}{dimension},
    xticklabel style={/pgf/number format/fixed,
                  /pgf/number format/precision=3},
    ytick={100,200,300,400,500,600,700,800},
    legend pos=north west,
    ymajorgrids=true,
    grid style=dashed,
    enlargelimits=false
]
\addplot[color=blue,mark=square] table[x=xlabel,y=triteration] {\triangleDimensionData};
\addlegendentry{Triangle}
\addplot[color=red,mark=x] table[x=xlabel,y=smoiteration] {\triangleDimensionData};
\addlegendentry{SMO}
\end{axis}
\end{tikzpicture}
}
}}
\subfloat[]{{
\resizebox {0.48\columnwidth} {!} {
\begin{tikzpicture}
\begin{axis}[
    title={Comparison of time triangle/SMO},
    xlabel={Dimensions},
    ylabel={Time(sec)},
    xmin=1, xmax=10,
    ymin=0, ymax=200,
    xtick= {1,2,3,4,5,6,7,8,9,10},
    xticklabels from table ={\dimensionlabel}{dimension},
    ytick={40,80,120,160,200},
    legend pos=north west,
    ymajorgrids=true,
    grid style=dashed,
    enlargelimits=false
]
 
\addplot[color=blue,mark=square] table[x=xlabel,y=trtime] {\triangleDimensionData};
\addlegendentry{Triangle}
\addplot[color=red,mark=x] table[x=xlabel,y=smotime] {\triangleDimensionData};
\addlegendentry{SMO}
\end{axis}
\end{tikzpicture}
}
}}
\caption{Performance of TA and SMO for calculation of optimal support} \label{fig:opsupport}
\end{figure}
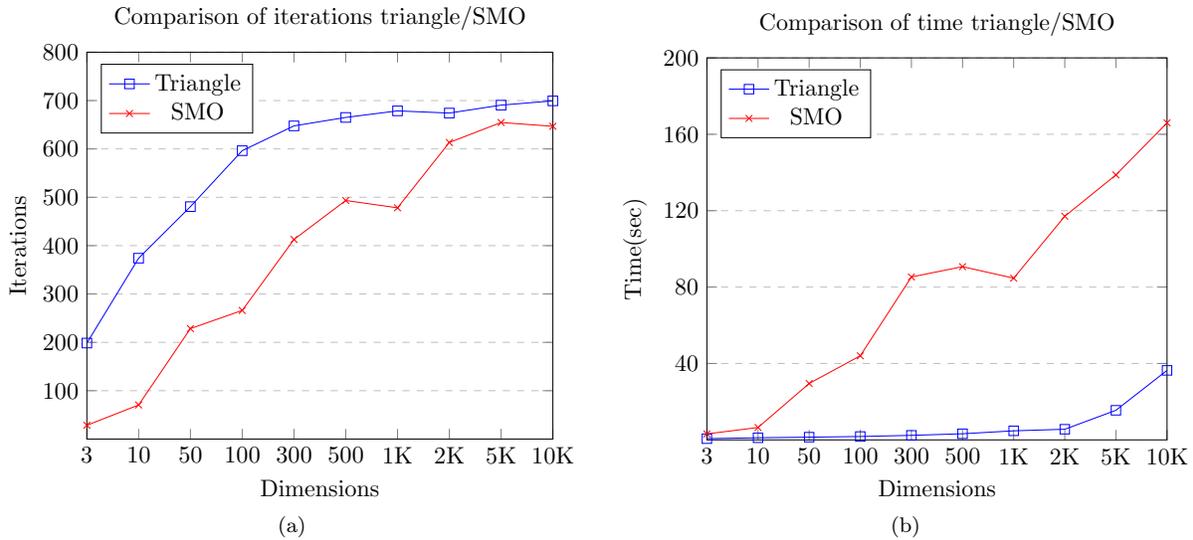
Based on our results we found that the Triangle Algorithm was faster in case of runtime against the SMO algorithm. This speed may be attributed to the linear work done in each iteration of the Triangle Algorithm, whereas each iteration of SMO is between linear and quadratic in, $n$, i.e. the number of points. 
We noticed that while being faster, the Triangle Algorithm calculates results similar to SMO in terms of sparsity and distance.

\subsection{Comparison Based on Distance}
In this experimental setup we generated points in $V=\{v_1 \dots v_n\}$ and $V'=\{v_1' \dots v_{n'}'\}$ from two unit balls. Let $K=conv(V)$ and $K'=conv(V')$. We translated the smaller ball along a random direction by $(1-k)max\{diam(K),diam(K')\},\quad k \in [0,1]$ units. Where diam is defined in (\ref{eq:12}). By adjusting the value of $k$ we were able to vary the distance between the convex hulls. The number of points in $V$ and $V'$ were 5000. The dimensionality was 1000.
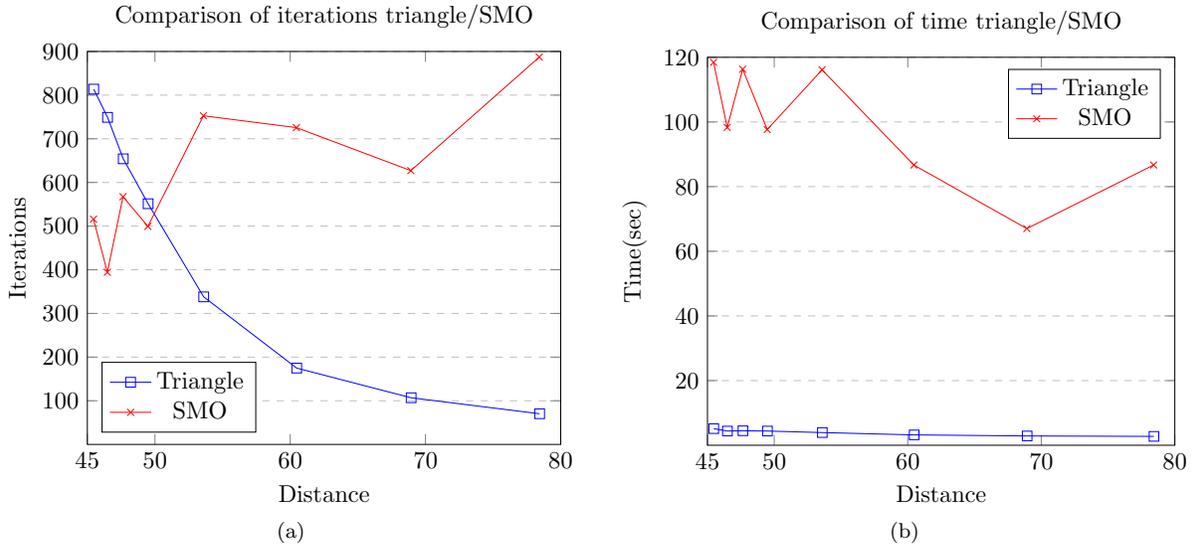
\begin{table}[H]
\centering
\pgfkeys{/pgf/number format/.cd, precision=3}
\pgfplotstabletypeset[
    %columns={dimension,triteration,trtime,trsparsity,trdist,smoiteration,smotime,smosparsity,smodist},
    columns={triteration,trtime,trdist,smoiteration,smotime,smodist},
    every head row/.style={%
        before row={\toprule%
            &\multicolumn{2}{c}{Triangle}  
            &\multicolumn{3}{c}{SMO} \\
            },
        after row=\hline
    },
    columns/triteration/.style={column name=Iterations, column type={|l}},
    columns/trtime/.style={column name=Time(sec), column type={|l}},
    %columns/trsparsity/.style={column name=Sparsity, column type={|l}},
    columns/trdist/.style={column name=Distance, column type={|l}},
    columns/smoiteration/.style={column name=Iterations, column type={|l}},
    columns/smotime/.style={column name=Time(sec), column type={|l}},
    %columns/smosparsity/.style={column name=Sparsity, column type={|l}},
    columns/smodist/.style={column name=Distance, column type={|l|}},
    every last row/.style={after row=\hline},
]\triangleDistanceData
\caption{Performance based on distance. Dimensions = 1000}
\end{table}
\begin{figure}[H]
\subfloat[]{{
\resizebox {0.48\columnwidth} {!} {
\begin{tikzpicture}
\begin{axis}[
    title={Comparison of iterations triangle/SMO},
    xlabel={Distance},
    ylabel={Iterations},
    xmin=45, xmax=80,
    ymin=0, ymax=900,
    xtick= {45,50,60,70,80},
    %xticklabels from table ={\triangleDistanceData}{trdist},
    ytick={100,200,300,400,500,600,700,800,900},
    legend pos=south west,
    ymajorgrids=true,
    grid style=dashed,
    enlargelimits=false
]
 
\addplot[color=blue,mark=square] table[x=trdist,y=triteration] {\triangleDistanceData};
\addlegendentry{Triangle}
\addplot[color=red,mark=x] table[x=smodist,y=smoiteration] {\triangleDistanceData};
\addlegendentry{SMO}
\end{axis}
\end{tikzpicture}
}
}}
\subfloat[]{{
\resizebox {0.48\columnwidth} {!} {
\begin{tikzpicture}
\begin{axis}[
    title={Comparison of time triangle/SMO},
    xlabel={Distance},
    ylabel={Time(sec)},
    xmin=45, xmax=80,
    ymin=0, ymax=120,
    xtick= {45,50,60,70,80},
    %xticklabels from table ={\triangleDistanceData}{trdist},
    ytick={20,40,60,80,100,120},
    legend pos=north east,
    ymajorgrids=true,
    grid style=dashed,
    enlargelimits=false
]
 
\addplot[color=blue,mark=square] table[x=trdist,y=trtime] {\triangleDistanceData};
\addlegendentry{Triangle}
\addplot[color=red,mark=x] table[x=smodist,y=smotime] {\triangleDistanceData};
\addlegendentry{SMO}
\end{axis}
\end{tikzpicture}
}
}}
\end{figure}

\section{Implementation Notes} \label{section:implementation}
\subsection{Closest Points Between Two Line Segments}
In {\it Triangle Algorithm I and II}, in Step 2 and 3 we either move p close to $p'$ along the segment $pv$ or moves  $p'$ close to $p$ along the segment $p'v'$. We can also jointly move $p, p'$ closer to each other by finding the points closest to each other along the line segments $pv$ and $p'v'$.
Given two lines $L_1 = p-v$ and $L_2 = p'-v'$, we wish to determine points $q, q'$ on $L_1$ and $L_2$ respectively such that 
\begin{equation}
    d(q,q') = min\{d(x,y): \quad x \in L_1, y \in L_2\}
\end{equation}

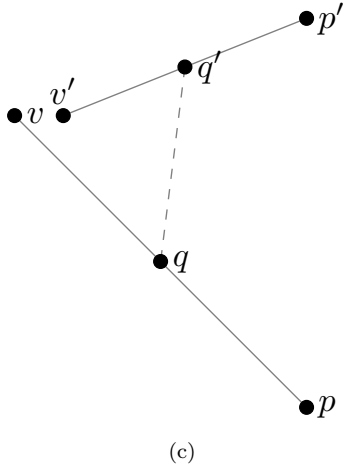
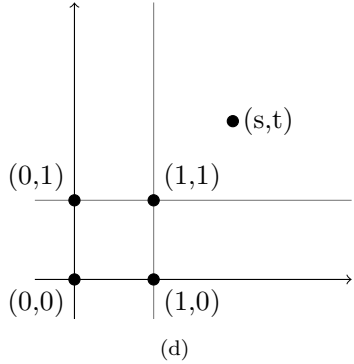
\begin{figure}[H]
\subfloat[]{{
\resizebox {0.3\columnwidth} {!} {
\begin{tikzpicture}
\draw[gray, thin] (1,4) -- (4,1);
\draw[gray, thin] (1.5,4) -- (4,5);
\draw[dashed, gray, thin] (2.75,4.5) -- (2.5,2.5);
\filldraw[black] (1,4) circle (2pt) node[anchor=west] {$v$}; 
\filldraw[black] (4,1) circle (2pt) node[anchor=west] {$p$}; 
\filldraw[black] (1.5,4) circle (2pt) node[anchor=south] {$v'$}; 
\filldraw[black] (4,5) circle (2pt) node[anchor=west] {$p'$}; 
\filldraw[black] (2.75,4.5) circle (2pt) node[anchor=west] {$q'$}; 
\filldraw[black] (2.5,2.5) circle (2pt) node[anchor=west] {$q$}; 
\end{tikzpicture}
}}
} \qquad \qquad \qquad
\subfloat[]{{
\resizebox {0.3\columnwidth} {!} {
\begin{tikzpicture}
\draw[-> ] (-0.5,0) -- (3.5,0);
\draw[->] (0,-0.5) -- (0,3.5);
\draw[gray, thin] (-0.5,1) -- (3.5,1);
\draw[gray, thin] (1,-0.5) -- (1,3.5);
\filldraw[black] (2,2) circle (2pt) node[anchor=west] {(s,t)};
\filldraw[black] (1,1) circle (2pt) node[anchor=south west] {(1,1)}; 
\filldraw[black] (0,1) circle (2pt) node[anchor=south east] {(0,1)}; 
\filldraw[black] (1,0) circle (2pt) node[anchor=north west] {(1,0)}; 
\filldraw[black] (0,0) circle (2pt) node[anchor=north east] {(0,0)}; 
\end{tikzpicture}
}}
}
\end{figure}

If $L_1$ and $L_2$ are not parallel and do not intersect, then the line segment $L_3 = qq'$ is perpendicular to both $L_1$ and $L_2$.

Let $\vec{a} = \overrightarrow{(v-p)}$ and $\vec{b} = \overrightarrow{(v'-p')}$. 
such that,
\begin{equation} \label{eq:15}
    q = p + s\vec{a}, \quad s \in [0,1]
    \qquad
    q' = p' + t\vec{b}, \quad t \in [0,1]
\end{equation}

then,
\begin{equation} \label{eq:17}
%\begin{split}
    L_3 = q-q' 
    \qquad
    L_3 = p + s\vec{a} - p' - t\vec{b}
%\end{split}
\end{equation}
Substituting (\ref{eq:17}) in (\ref{eq:15}), we get
\begin{equation}
    (\vec{a}.\vec{a})s - (\vec{a}.\vec{b})t = -\vec{a}.(p-p')
    \qquad
    (\vec{b}.\vec{a})s - (\vec{b}.\vec{b})t = -\vec{b}.(p-p')
\end{equation}

Solving for s and t, we get

\begin{equation} \label{eq:20}
    s=\frac{(\vec{a}.\vec{b})(\vec{b}.(p-p')) - (\vec{b}.\vec{b})(\vec{a}.(p-p'))}{(\vec{a}.\vec{a})(\vec{b}.\vec{b}) - (\vec{a}.\vec{b})^2}
\end{equation}

\begin{equation} \label{eq:21}
    t=\frac{(\vec{a}.\vec{a})(\vec{b}.(p-p')) - (\vec{a}.\vec{b})(\vec{a}.(p-p'))}{(\vec{a}.\vec{a})(\vec{b}.\vec{b}) - (\vec{a}.\vec{b})^2}
\end{equation}

If the lines are parallel the denominator in (\ref{eq:20}) and (\ref{eq:21}) goes to zero. In this case we can keep one of the parameters constant, say s=0, and solve for t.

After getting the initial estimates for $(s,t)$ we wish to clip them such that the point lies between the line segments $[v,p]$ and $[v',p']$. (\ref{eq:15})

\begin{equation}
s = 
\begin{cases}
    0, & \text{if } s < 0 \\
    s, & \text{if } 0 \le s \le 1 \\
    1, & \text{if } s > 1
\end{cases}
\qquad
t = 
\begin{cases}
    0, & \text{if } t < 0 \\
    t, & \text{if } 0 \le t \le 1 \\
    1, & \text{if } t > 1
\end{cases}
\end{equation}

This method using the dot product works for points in any number of dimensions.~\cite{linetoline}

\subsection{Caching Results}
Step 2 and 3 of the {\it Triangle Algorithm I} , solves a maximization problem for finding the pivots i.e. if the value of $max\{(p-p')^Tv : \quad v \in K\}) < \frac{1}{2} (||p'||^2 - ||p||^2)$ then v acts as a pivot.

Notice that we solve the same problem in Step 3 and 4 of {\it Triangle Algorithm II}. 
We can cache these results from Triangle Algorithm I and use it in Triangle Algorithm II. 
Specifically
\linebreak

\RestyleAlgo{boxruled}
\resizebox{0.95\columnwidth}{!}{
\LinesNumbered
\begin{algorithm}[H]
\SetAlgoLined
  \eIf{$max\{(p-p')^Tv : \quad v \in K\}) < \frac{1}{2} (||p'||^2 - ||p||^2)$}{
     $pivot \gets v$\;
  }{
     $extreme point \gets v$\;
  }
\caption{Pivot selection}
\end{algorithm}
}
\linebreak
During each iteration we update the iterates as
\begin{equation}
\begin{split}
p^{new} = (1-\alpha)p^{old} + \alpha v \\
p'^{new} = (1-\alpha ')p'^{old} + \alpha' v'
\end{split}
\end{equation}
 For a vector $x$ let,
\begin{equation}
E_p = p^Tx \quad {E_{p'}} = {p'}^Tx
\end{equation}
therefore,
\begin{equation}
E = (p-p')^Tx = E_p - {E_{p'}}
\end{equation}
 We can update the errors as
    \begin{equation}
    \begin{split}
    {E_p}^{new} = (1-\alpha){E_p}^{old} + \alpha v^Tx \\
    {E_{p'}}^{new} = (1-\alpha'){E_{p'}}^{old} + \alpha' {v'}^Tx
    \end{split}
    \end{equation}
The update for the error $E$ is 
{
    \begin{equation}
        E^{new} = {E_p}^{new} - {E_{p'}}^{new}
    \end{equation}
}
If we precompute the dot-products, similar to SMO, each update is $O(n)$.

\subsection{Avoiding Zig-Zag}
The Triangle Algorithm has been observed to suffer from zig-zagging. In this section we discuss the case for the same and a simple idea to overcome it.
Let $V=\{v_1,\dots,v_n\}$, $K=conv(V)$, $p \in K$ and we wish to move $p$ closer to $p'$ 
\begin{figure}
\centering
\definecolor{zzttqq}{rgb}{0.6,0.2,0.}
\definecolor{sqsqsq}{rgb}{0.12549019607843137,0.12549019607843137,0.12549019607843137}
\begin{tikzpicture}[line cap=round,line join=round,>=triangle 45,x=1.0cm,y=1.0cm,scale=0.8]
\clip(11.038958775710709,10.579781573835277) rectangle (26.499873816514402,18.285156493586907);
\draw [dash pattern=on 1pt off 1pt] (15.062207920008118,16.429044045469652)-- (22.47360687532242,16.495966429420566);
\draw [dash pattern=on 1pt off 1pt on 2pt off 4pt] (18.56976615613743,11.334512592645765)-- (18.543565882854306,16.94137107523362);
\draw [line width=0.4pt,dash pattern=on 1pt off 1pt] (18.56976615613743,11.334512592645765)-- (15.062207920008118,16.429044045469652);
\draw [line width=0.4pt,dash pattern=on 1pt off 1pt] (18.56976615613743,11.334512592645765)-- (22.47360687532242,16.495966429420566);
\draw (18.543565882854306,16.94137107523362)-- (16.01914419215385,15.039147824623345);
\draw [dash pattern=on 1pt off 1pt on 2pt off 4pt] (16.01914419215385,15.039147824623345)-- (22.47360687532242,16.495966429420566);
\draw (18.543565882854306,16.94137107523362)-- (18.917211868984914,15.69326252274975);
\draw [dash pattern=on 1pt off 1pt on 2pt off 4pt] (15.062207920008118,16.429044045469652)-- (18.917211868984914,15.69326252274975);
\draw (18.543565882854306,16.94137107523362)-- (18.004195332067113,15.867524522657723);
\draw [dash pattern=on 1pt off 1pt on 2pt off 4pt] (18.004195332067113,15.867524522657723)-- (22.47360687532242,16.495966429420566);
\draw (18.543565882854306,16.94137107523362)-- (18.70891540742369,15.966614889164578);
\draw [dash pattern=on 1pt off 1pt on 2pt off 4pt] (18.70891540742369,15.966614889164578)-- (15.062207920008118,16.429044045469652);
\draw (18.543565882854306,16.94137107523362)-- (18.362900339883247,16.010492128099628);
\draw (14.0,16.63973789093161) node[anchor=north west] {$\mathbf{v_1}$};
\draw (22.617087235858314,16.69993613249217) node[anchor=north west] {$\mathbf{v_2}$};
\draw (18.67410241364167,17.301918548097767) node[anchor=north west] {$\mathbf{p'}$};
\draw (18.704201534421948,11.532920398544137) node[anchor=north west] {$\mathbf{p_1}$};
\begin{scriptsize}
\draw [fill=black] (15.062207920008118,16.429044045469652) circle (2.5pt);
\draw [fill=black] (22.47360687532242,16.495966429420566) circle (2.5pt);
\draw [fill=sqsqsq] (18.543565882854306,16.94137107523362) circle (2.5pt);
\draw [fill=zzttqq] (18.56976615613743,11.334512592645765) circle (2.5pt);
\draw [fill=zzttqq] (16.01914419215385,15.039147824623345) circle (2.5pt);
\draw [fill=zzttqq] (18.917211868984914,15.69326252274975) circle (2.5pt);
\draw [fill=zzttqq] (18.004195332067113,15.867524522657723) circle (2.5pt);
\draw [fill=zzttqq] (18.70891540742369,15.966614889164578) circle (2.5pt);
\draw [fill=zzttqq] (18.362900339883247,16.010492128099628) circle (2.5pt);
\end{scriptsize}
\end{tikzpicture}
\caption{Case for Zig Zag.} \label{FigZigZag}
\end{figure}
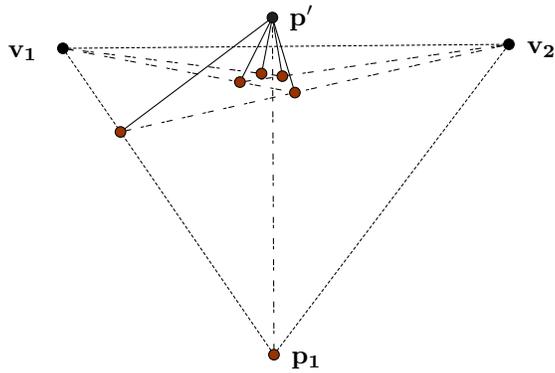
In this case both $v_1$ and $v_2$ are suitable for pivot selection. But $p$ can either move along the line segment $pv_1$ or $pv_2$, and this causes it to zig-zag between $v_1$ and $v_2$ by choosing them as a pivot in an alternating fashion. We wish to move it along the segment $pp'$ instead. 
Figure \ref{FigZigZag} shows the trajectory of the movement of the iterate that is zig-zagging between $v_1$ and $v_2$.
A simple solution would be to add another point $v_{ext} = (v_1 + v_2)/2$ which will be a better pivot choice and move $p$ along the desired trajectory. 
To identify zig-zagging we can check if $\triangle d(p,p') < \epsilon d(p,p')$ where $\epsilon \in [0,1]$. If this condition is met we can add the midpoint of two most frequently used pivots. This approach has been found to be working well in our experiments.

\subsection{Reducing Floating Point Operations}
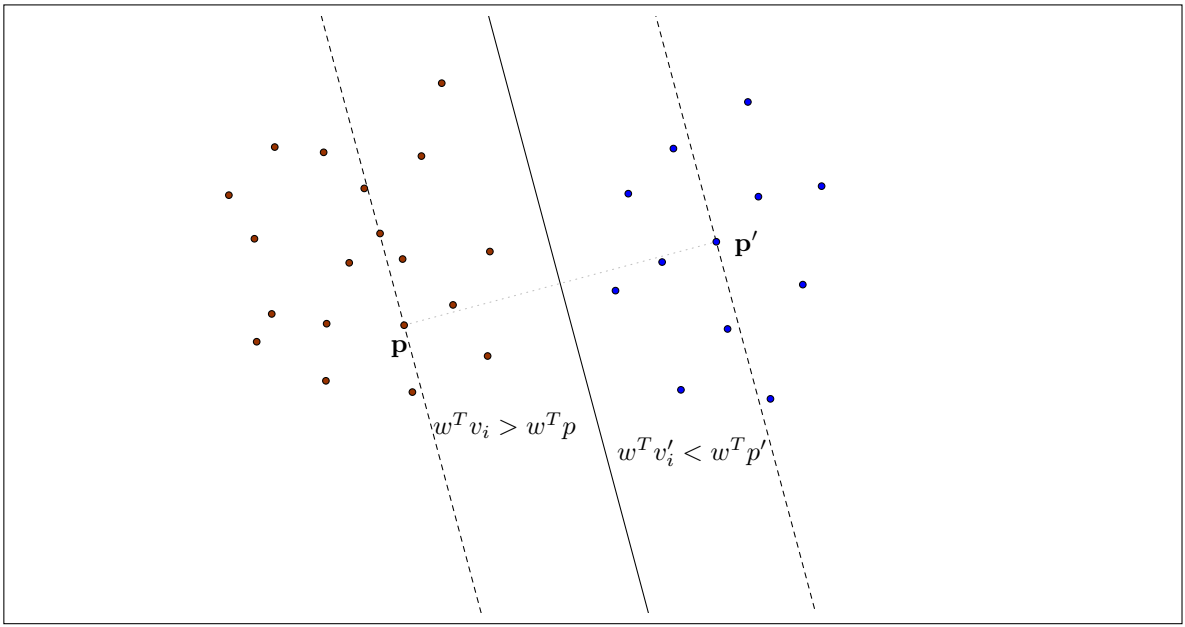
\begin{figure}[H]
    \centering
    \begin{tikzpicture}[line cap=round,line join=round,>=triangle 45,x=1.0cm,y=1.0cm,scale=0.5,framed]
\clip(-4.66,-6.76) rectangle (26.06,9.1);
\draw [domain=-4.66:26.06] plot(\x,{(-86.0068--8.3*\x)/-2.22});
\draw [dash pattern=on 2pt off 2pt,domain=-4.66:26.06] plot(\x,{(-49.0976--8.3*\x)/-2.22});
\draw [dash pattern=on 2pt off 2pt,domain=-4.66:26.06] plot(\x,{(-122.916--8.3*\x)/-2.22});
\draw (5.08,0.72) node[anchor=north west] {$\mathbf{p}$};
\draw (14.22,3.64) node[anchor=north west] {$\mathbf{p'}$};
\draw [dotted,color=cqcqcq] (5.68,0.88)-- (13.98,3.1);
\draw (6.2,-1.18) node[anchor=north west] {$w^Tv_i > w^Tp$};
\draw (11.1,-1.86) node[anchor=north west] {$w^Tv_i' < w^Tp'$};
\begin{scriptsize}
\draw [fill=qqqqff] (11.64,4.38) circle (2.5pt);
\draw [fill=qqqqff] (12.54,2.56) circle (2.5pt);
\draw [fill=qqqqff] (15.1,4.3) circle (2.5pt);
\draw [fill=qqqqff] (12.84,5.58) circle (2.5pt);
\draw [fill=qqqqff] (13.98,3.1) circle (2.5pt);
\draw [fill=qqqqff] (16.28,1.96) circle (2.5pt);
\draw [fill=qqqqff] (16.78,4.58) circle (2.5pt);
\draw [fill=qqqqff] (14.82,6.82) circle (2.5pt);
\draw [fill=qqqqff] (15.42,-1.08) circle (2.5pt);
\draw [fill=qqqqff] (13.04,-0.84) circle (2.5pt);
\draw [fill=qqqqff] (14.28,0.78) circle (2.5pt);
\draw [fill=qqqqff] (11.3,1.8) circle (2.5pt);
\draw [fill=zzttqq] (5.04,3.32) circle (2.5pt);
\draw [fill=zzttqq] (4.22,2.54) circle (2.5pt);
\draw [fill=zzttqq] (5.64,2.64) circle (2.5pt);
\draw [fill=zzttqq] (6.14,5.38) circle (2.5pt);
\draw [fill=zzttqq] (3.54,5.48) circle (2.5pt);
\draw [fill=zzttqq] (1.7,3.18) circle (2.5pt);
\draw [fill=zzttqq] (1.76,0.44) circle (2.5pt);
\draw [fill=zzttqq] (3.6,-0.6) circle (2.5pt);
\draw [fill=zzttqq] (5.9,-0.9) circle (2.5pt);
\draw [fill=zzttqq] (5.68,0.88) circle (2.5pt);
\draw [fill=zzttqq] (7.9,0.06) circle (2.5pt);
\draw [fill=zzttqq] (7.96,2.84) circle (2.5pt);
\draw [fill=zzttqq] (6.98,1.42) circle (2.5pt);
\draw [fill=zzttqq] (3.62,0.92) circle (2.5pt);
\draw [fill=zzttqq] (1.02,4.34) circle (2.5pt);
\draw [fill=zzttqq] (2.16,1.18) circle (2.5pt);
\draw [fill=zzttqq] (4.62,4.52) circle (2.5pt);
\draw [fill=zzttqq] (2.24,5.62) circle (2.5pt);
\draw [fill=zzttqq] (6.68,7.32) circle (2.5pt);
\end{scriptsize}
\end{tikzpicture}
    \caption{Non bounding points}
    \label{fig:nonbounding}
\end{figure}

A remarkable property of Triangle Algorithm is that it continuously reduces the gap between the iterates $p \in conv(V)$ and $p' \in conv(V')$ in each iteration. This gap $d(p,p')$ also gives an estimate of the distance between the convex hulls. Between each iterations we have some points that are non-bounding i.e. $w^Tv_i > w^Tp, \quad \forall v_i \in V$ and $w^Tv_i' < w^Tp', \quad \forall v_i' \in V'$. Therefore, it makes sense to search for the next-pivot or extreme points within these non-bounding points first. 
Figure \ref{fig:nonbounding} shows the non-bounding points where $p$ and $p'$ act as iterates.

If none of them satisfy the property then we can scan all the points. Notice that when we have the optimal hyperplane then we will find no such points in both the non-bounding set and the complete set.
This is a heuristic idea to reduce the number of scans over the whole input.

\section{Future Work}
Our experiments show that the Triangle Algorithm does well on the hard-margin formulation of the optimization problem. In this section we present the idea for solving the soft-margin problem using the Triangle Algorithm.

We use the following formulation for the soft-margin problem
\begin{equation}
\begin{split}
    & min \quad \frac{1}{2} \Vert w \Vert^2 + C\sum_{i=1}^n \xi_i^2 \\
    & s.t.\quad y_i(w^Tx^{(i)} + b) \ge 1 - \xi_i, \quad y_i \in \{-1,1\}
\end{split}
\end{equation}

If we let
\begin{equation}
\eta_i = \sqrt{\frac{C}{2}}\xi_i, \quad i=1,\dots,n
\end{equation}
and use the following problem,
\begin{equation}
\begin{split}
    & min \quad \frac{1}{2} \Vert w \Vert^2 +  \frac{1}{2} \Vert \eta \Vert^2\\
    & s.t.\quad y_i(w^Tx^{(i)} + b) + \sqrt{\frac{2}{c}}\eta_i \ge 1, \quad y_i \in \{-1,1\}
\end{split}
\end{equation}
it is solvable by the Triangle Algorithm.

\bibliography{egbib}
\bibliographystyle{ieeetr}

\end{document}